\documentclass{article}
\usepackage{graphicx}
\usepackage{fancyhdr}

\fancypagestyle{firstpage}{
  \fancyhf{}
  \fancyhead[L]{\includegraphics[width=3.8cm]{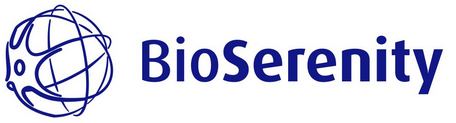}}
  
}

\usepackage[english]{babel}

\usepackage[a4paper,top=2cm,bottom=2cm,left=1.3cm,right=1.3cm,marginparwidth=1.75cm]{geometry}

\usepackage{amsmath}
\usepackage{graphicx}
\graphicspath{{figures/}}
\usepackage[colorlinks=true, allcolors=blue]{hyperref}
\usepackage[utf8]{inputenc}
\usepackage[noblocks]{authblk}
\usepackage{float}
\usepackage[toc,page]{appendix}
\usepackage{lmodern} 
\usepackage{hyperref}

\usepackage{multirow}
\usepackage{booktabs}
\usepackage[table,xcdraw,dvipsnames]{xcolor}

\definecolor{DarkBlue}{rgb}{0.0, 0.1, 0.6}

\hypersetup{
    colorlinks=true,
    citecolor=OliveGreen,  
    linkcolor=black, 
    urlcolor=green   
}

\usepackage{caption}
\title{\textbf{\fontsize{20}{32}\selectfont{BioSerenity-E1}: a self-supervised EEG model for medical applications }\vspace{0.5cm}}

\author[1\(\dagger\)\(\ast\)]{Ruggero G. Bettinardi}
\author[1\(\dagger\)]{Mohamed Rahmouni}
\author[1]{Ulysse Gimenez}

\affil[1]{\textbf{Data Science Team @ BioSerenity}}

\date{}

\begin{document}
\maketitle\thispagestyle{firstpage}

{\renewcommand{\thefootnote}{}\footnote{\(\dagger\) These authors contributed equally}}
{\renewcommand{\thefootnote}{}\footnote{\(\ast\) Corresponding author (ruggero.bettinardi@bioserenity.com)}}

\begin{abstract}
Electroencephalography (EEG) serves as an essential diagnostic tool in neurology; however, its accurate manual interpretation is a time-intensive process that demands highly specialized expertise, which remains relatively scarce and not consistently accessible. To address these limitations, the implementation of automated pre-screening and analysis systems for EEG data holds considerable promise. Traditional automated approaches relying on handcrafted features struggle to capture the complex spatiotemporal patterns in EEG signals, particularly given their low signal-to-noise ratio and inherent biological variability, whereas more performant end-to-end deep learning architectures are dependent on large labeled datasets that are difficult and costly to acquire, especially in medical contexts. Advances in self-supervised learning made it possible to pre-train complex deep learning architectures on large volumes of unlabeled EEG data to learn generalizable representations, that can later be used to enhance performance on multiple tasks while needing less downstream data. In the present paper, we introduce \textcolor{DarkBlue}{BioSerenity-E1}, the first of a family of self-supervised foundation models for clinical EEG applications that combines spectral tokenization with masked prediction to achieve state-of-the-art performance across relevant diagnostic tasks. The two-phase self-supervised pretraining framework initially acquires compressed EEG representations via a transformer-based VQ-VAE architecture designed to reconstruct log-multitaper spectral projections, then implements extensive (70\% block) masked token prediction to force the model to learn complex spatiotemporal dependencies in EEG signals. \textcolor{DarkBlue}{BioSerenity-E1} achieves strong performance across three clinical tasks, either in line or above state-of-the-art methods: seizure detection (AUROC = 0.926 ± 0.002, Sensitivity =  0.909 ± 0.035), normal/abnormal classification (AUPRC = 0.970 ± 0.001 on proprietary data; 0.910 ± 0.002 on TUH-Abnormal), and multiclass pathology differentiation on unbalanced data (Weighted F1 = 0.730 ± 0.001). The utility of \textcolor{DarkBlue}{BioSerenity-E1} is further confirmed in low-data regimes scenarios,  showing clear improvements in AUPRC (from +2\% to 17\%) when trained on less than 10\% of the available data.\vspace{1cm}
\end{abstract}

\section{Introduction}

Electroencephalography (EEG) is a non-invasive technique that measures and records the brain's electrical activity through electrodes placed on the scalp. This technology captures the electrical signals generated by neurons communicating via synaptic excitations, providing real-time insights into brain function \cite{silipo1998dynamics, cabral2014exploring, blinowska2006}. This has made EEG a valuable source of information enabling applications in multiple domains. In the medical field, EEG has been shown to provide detailed and relevant information in several health conditions \cite{bera2021}, such as epilepsy and seizure disorders \cite{noachtar2022}, sleep disorders \cite{behzad2021role}, neurodegenerative diseases as Alzheimer, Parkinson and other forms of dementia, traumatic brain injuries and psychiatric conditions \cite{jadhav2022, cruzat2023temporal,dattola2024}. Impressive advances have also been made using EEG to develop brain-computer interfaces (BCI, \cite{lotte2015, peksa2023}), devices that leverage the brain’s  electrical potentials recorded at the level of the scalp to guide and accelerate rehabilitation for patients affected by stroke or spinal cord injuries \cite{lazarou2018}, wheelchair control for paralyzed individuals \cite{zhang2023recent}, to build communication systems for locked-in patients \cite{machado2010eeg}, as well as a plethora of non-medical applications ranging from emotion recognition \cite{de2021classification, jafari2023emotion} and cognitive state monitoring \cite{panwar2024eeg} to entertainment and gaming interfaces \cite{liao2012gaming}.\vspace{0.25cm} 

Early attempts at automatize the interpretation of the EEG signal relied on manually defining signal features later used to perform analyses and classification of the EEG \cite{koles1991quantitative, schreiter1994quantitative, nuwer1996quantitative, da1999eeg, coburn2006value, song2015background}. However, the identification and definition of these features is a non-trivial task even for experienced professionals. The EEG is in fact a very complex type of signal emerging by the simultaneous activity and interplay of tens of thousands of neurons at different spatial and temporal scales, characterized by periodic and aperiodic components, large inter-subject and intra-subject variability and multiple sources of artifacts and interference, leading to poor signal-to-noise ratio \cite{blanco1995stationarity, kondacs1999long, dustman1999life, lopes2000rhythms, bettinardi2016spontaneous}.\vspace{0.25cm} 
 
Modern analysis approaches employ various deep learning architectures, as they enable end-to-end processing of EEG signals, eliminating the need for manual feature extraction. This provide several, ground-breaking advantages over other approaches: direct learning from raw EEG data, improving model performance, automatic detection of hidden, meaningful temporal and spatial patterns in noisy signals that might be missed by other methods \cite{roy2019deep, craik2019deep, stieger2021, kalita2024aneeg, wang2024explain}. Example of successful deep learning networks applied to EEG include Convolutional Neural Networks (CNN, \cite{lawhern2018eegnet}), Long Short-Term Memory (LSTM) networks \cite{alhagry2017emotion}, Graph Neural Networks (GNN, \cite{tang2021self}), Transformer models \cite{song2021transformer, song2022eeg, abibullaev2023deep} as well as hybrid architectures obtained combined different types of architectures \cite{craley2021automated, zhang2023spatial}.\vspace{0.25cm} 

\textbf{Related Work}. The field of EEG foundation modeling has evolved significantly since 2021, driven by advancements in self-supervised learning (SSL) and transformer architectures. BENDR pioneered this domain by adapting language-modeling techniques to EEG data, enabling cross-hardware compatibility and task adaptability through fine-tuning \cite{kostas2021bendr}. Building on this, MAEEG introduced masked auto-encoding with transformers, demonstrating that reconstructing larger masked EEG portions improved sleep stage classification accuracy under limited labeled data scenarios \cite{chien2022maeeg}. Subsequent innovations saw BrainBERT applying transformer architectures to intracranial recordings, showing that unsupervised pretraining reduced data requirements for neural decoding tasks \cite{wang2023brainbert}, while BIOT addressed biosignal heterogeneity through unified tokenization with channel/position embeddings, facilitating cross-modal learning \cite{biot2023}, and MBrain incorporated graph neural networks to model spatial brain correlations \cite{cai2023mbrain}. Recent efforts have focused on scaling and specialization: NeuroGPT integrated GPT architectures with EEG encoders to enhance motor imagery classification in low-data regimes \cite{cui2024neuro}, whereas EEGFormer emerged as the first interpretable foundation model with inherent anomaly detection capabilities \cite{chen2024eegformer}. Brant-2 extended intracranial modeling to broader neural data types while maintaining performance with scarce labels \cite{yuan2024brant}, and LaBraM achieved cross-dataset compatibility through neural tokenization, pretrained on over 2,500 EEG hours across 20 datasets \cite{labram2024}. FoME introduced adaptive temporal-spectral attention scaling \cite{shi2024fome}. Masking strategies advanced with EEG2Rep employing latent-space prediction and semantic-preserving masking to enhance noise robustness \cite{mohammadi2024eeg2rep}, while BrainWave scaled to 40,000+ hours of multimodal neural data \cite{yuan2024brainwave}. Lastly, architectural innovations include Graph-Enhanced models combining GNNs with masked autoencoders for spatiotemporal modeling \cite{wang2024graph}, utilizing criss-cross transformers with conditional positional encoding as in CBraMod \cite{wang2024cbramod}, and even integrating tokenized EEG signals into a large language model (LLM) that learns causal EEG information via multi-channel autoregression \cite{jiang2025neurolm}.\vspace{0.25cm} 

\textbf{BioSerenity-E1} is an EEG foundation model pre-trained on 4000 hours of EEG obtained from clinical settings. The pre-training is based on self-supervised learning and is performed in two phases: an EEG tokenizer model is first tasked to learn a compress, discrete latent representation of the spectral features characterizing the EEG signal by reconstructing its power spectrum; in the second phase, a twin deep transformer network is trained to learn to predict the latent representations associated to partially masked input EEG signals. This two-stage pre-training strategy, inspired by \cite{labram2024}, effectively pushes the model to learn generalizable features of the input space relying on both local and global relationships between different channels and temporal segments. The choice of reconstructing the power spectrum instead of the raw EEG signal as a proxy task to learn latent representations is due to the fact that raw EEG is inherently characterized by a low signal-to-noise ratio, while its spectral representation tends to be more stable over time, making it a more reliable target for self-supervised learning \cite{wu2024neuro}.\vspace{0.25cm}  

\textbf{Contributions}. \textcolor{DarkBlue}{BioSerenity-E1} builds upon recent advances in the field of EEG deep representation learning while incorporating several modifications. We estimate the power spectral distribution of the input signal using multitaper discrete prolate spheroidal sequences \cite{thomson1982spectrum, press2007numerical}, a method well-suited to effectively suppress the influence of non-stationarities and artifacts commonly encountered in electrophysiological data, while providing statistically robust spectral estimates \cite{van2007comparison, melman2016robust}. In the EEG tokenization phase, we employ the logarithm of the estimated power spectrum distribution to minimize reconstruction loss, as it rescales the power distribution in a way that enhances differences across frequencies in a range (1 to 45 Hz) meaningful for EEG analysis in clinical contexts \cite{tatum2021handbook}. Additionally, we improved the masked-token prediction strategy by using multiple large masks and, at the same time, masking consistent portions (70\%) of the input, while calculating the prediction loss on both masked and unmasked patches to improve overall optimization. All these modifications altogether make training more challenging, further forcing the model to learn relevant latent features characterizing the EEG signal. Finally, our model is implemented using BF16 precision to optimize computational efficiency without compromising accuracy.\vspace{0.25cm}

\section{Methods}\vspace{0.25cm}

\subsection{Datasets and Preprocessing}\vspace{0.25cm}

\textbf{Pretraining Datasets.} 
We combined EEG records from two proprietary databases and four public datasets belonging to the TUH EEG Corpus, a well-known public database of clinical EEG records \cite{harati2014tuh}, to build a pre-training dataset of 4000 EEG hours (see Table~\ref{tab:pretrain-ds}). \textit{Bioserenity-Neurophy-FR1} is a database that includes both clinically normal or altered EEG (sedation, epilepsy, encephalopathy, lesion, etc), recorded with either Micromed or BioSerenity’s Neuronaute EEG system, obtained in BioSerenity centers, ICUs, hospitals and private clinics in France from 2021 to 2024 (45\% female patients, mean age 63±22 years) . \textit{Bioserenity-US1} comprises fully anonymized long continuous EEG (24 to 75 hours) recorded using the Compumedics EEG system from US patients under epilepsy monitoring from 2021 to 2023. The four TUH datasets we used to create the pretraining dataset were the “train” subsets of \textit{TUH-Abnormal} \cite{lopez2015automated}, \textit{TUH-Seizure} \cite{shah2018temple}, \textit{TUH-Events} \cite{harati2015improved} and \textit{TUH-Artifact} \cite{hamid2020temple}. We set 6 years old as minimal age and record duration of at least 5 minutes as inclusion criteria. Records selected to build the pre-training datasets listed in Table~\ref{tab:pretrain-ds} all belonged to the corresponding “train” subsets: no records from the “test” sets were included in pre-training \textcolor{DarkBlue}{BioSerenity-E1}.\vspace{0.35cm}

\begin{table}[h]
\small
\centering
\renewcommand{\arraystretch}{1.2} 
\begin{tabular}{lcccc}
                         & \textbf{EEG hours} & \textbf{Windows} & \textbf{Tokens} & \textbf{Percentage} \\ \hline
Bioserenity-Neurophy-FR1 & 2,803              & 630, 684         & 161.4M          & 70\%                \\
Bioserenity-US1          & 1,201              & 270,336          & 69.2M           & 20\%                \\
TUH-Seizure              & 200                & 45,056           & 11.5M           & 5\%                 \\
TUH-Abnormal             & 160                & 36,044           & 9.2M            & 4\%                 \\
TUH-Events               & 28                 & 6,308            & 1.6M            & 0.7\%               \\
TUH-Artifacts            & 12                 & 2,703            & 0.7M            & 0.3\%               \\ \hline
\textbf{TOTAL}           & \textbf{4,005}     & \textbf{901,120} & \textbf{253.6M} & \textbf{100\%}     
\end{tabular}
\caption{\textbf{Pretraining dataset}}
\label{tab:pretrain-ds}
\end{table}

\textbf{Downstream Datasets.} The pretrained foundation models were tested on different downstream tasks from 4 fine-tuning datasets (see Table~\ref{tab:downstream-ds}). The continuous EEG signals of all records in each of these datasets were first divided into non-overlapping windows of 16 seconds (each storing the signal of 16 channels), and each window was then assigned with a unique class label to predict. \textit{Neurophy-Abnormal} and \textit{Neurophy-Multiclass} are a subset of \textit{Bioserenity-Neurophy-FR1} a proprietary clinical database composed by 338 hours of EEG from 7536 records reviewed by an expert neurologist and labelled as either normal or representative of three broad types of abnormality (lesion, status epilepticus, encephalopathy). \textit{TUH-Abnormal} is a balanced dataset containing 1006 hours of EEG from records classified as clinically normal or abnormal that was drawn from \cite{lopez2015automated}. \textit{TUH-Seizure} contains 138 hours of EEG from records with annotated seizure events and normal background activity drawn from \cite{shah2018temple}. Every window including a seizure event of at least 3 seconds was assigned the label "seizure".\vspace{0.35cm}

\begin{table}[t]
\small
\centering
\renewcommand{\arraystretch}{1.2} 
\begin{tabular}{cccccc}
\hline
\textbf{Dataset} &
  \textbf{Subset} &
  \textbf{EEG hours} &
  \textbf{Windows} &
  \textbf{Tokens} &
  \textbf{Classes (windows, \%)} \\ \hline
 &
  \cellcolor[HTML]{EFEFEF} &
  \cellcolor[HTML]{EFEFEF} &
  \cellcolor[HTML]{EFEFEF} &
  \cellcolor[HTML]{EFEFEF} &
  \cellcolor[HTML]{EFEFEF}Normal (31050, 51\%) \\
 &
  \multirow{-2}{*}{\cellcolor[HTML]{EFEFEF}train} &
  \multirow{-2}{*}{\cellcolor[HTML]{EFEFEF}270} &
  \multirow{-2}{*}{\cellcolor[HTML]{EFEFEF}60,850} &
  \multirow{-2}{*}{\cellcolor[HTML]{EFEFEF}15.5M} &
  \cellcolor[HTML]{EFEFEF}Abnormal (29800, 49\%) \\
 &
   &
   &
   &
   &
  Normal (7800, 51\%) \\
\multirow{-4}{*}{Neurophy-Abnormal} &
  \multirow{-2}{*}{test} &
  \multirow{-2}{*}{68} &
  \multirow{-2}{*}{15,285} &
  \multirow{-2}{*}{3.9M} &
  Abnormal (7485, 49\%) \\ \hline
 &
  \cellcolor[HTML]{EFEFEF} &
  \cellcolor[HTML]{EFEFEF} &
  \cellcolor[HTML]{EFEFEF} &
  \cellcolor[HTML]{EFEFEF} &
  \cellcolor[HTML]{EFEFEF}Normal (31050, 51\%) \\
 &
  \cellcolor[HTML]{EFEFEF} &
  \cellcolor[HTML]{EFEFEF} &
  \cellcolor[HTML]{EFEFEF} &
  \cellcolor[HTML]{EFEFEF} &
  \cellcolor[HTML]{EFEFEF}Lesion (11100, 18\%) \\
 &
  \cellcolor[HTML]{EFEFEF} &
  \cellcolor[HTML]{EFEFEF} &
  \cellcolor[HTML]{EFEFEF} &
  \cellcolor[HTML]{EFEFEF} &
  \cellcolor[HTML]{EFEFEF}Status (9670, 16\%) \\
 &
  \multirow{-4}{*}{\cellcolor[HTML]{EFEFEF}train} &
  \multirow{-4}{*}{\cellcolor[HTML]{EFEFEF}270} &
  \multirow{-4}{*}{\cellcolor[HTML]{EFEFEF}60,850} &
  \multirow{-4}{*}{\cellcolor[HTML]{EFEFEF}15.5M} &
  \cellcolor[HTML]{EFEFEF}Encephalopathy (9030, 14\%) \\
 &
   &
   &
   &
   &
  Normal (7800, 51\%) \\
 &
   &
   &
   &
   &
  Lesion (2730, 18\%) \\
 &
   &
   &
   &
   &
  Status (2415, 16\%) \\
\multirow{-8}{*}{Neurophy-Multiclass} &
  \multirow{-4}{*}{test} &
  \multirow{-4}{*}{68} &
  \multirow{-4}{*}{15,285} &
  \multirow{-4}{*}{3.9M} &
  Encephalopathy (2340, 14\%) \\ \hline
 &
  \cellcolor[HTML]{EFEFEF} &
  \cellcolor[HTML]{EFEFEF} &
  \cellcolor[HTML]{EFEFEF} &
  \cellcolor[HTML]{EFEFEF} &
  \cellcolor[HTML]{EFEFEF}Normal (101257, 49\%) \\
 &
  \multirow{-2}{*}{\cellcolor[HTML]{EFEFEF}train} &
  \multirow{-2}{*}{\cellcolor[HTML]{EFEFEF}904} &
  \multirow{-2}{*}{\cellcolor[HTML]{EFEFEF}203,388} &
  \multirow{-2}{*}{\cellcolor[HTML]{EFEFEF}52M} &
  \cellcolor[HTML]{EFEFEF}Abnormal (102131, 51\%) \\
 &
   &
   &
   &
   &
  Normal (12412, 54\%) \\
\multirow{-4}{*}{TUH-Abnormal} &
  \multirow{-2}{*}{test} &
  \multirow{-2}{*}{102} &
  \multirow{-2}{*}{23,040} &
  \multirow{-2}{*}{5.8M} &
  Abnormal (10628, 46\%) \\ \hline
 &
  \cellcolor[HTML]{EFEFEF} &
  \cellcolor[HTML]{EFEFEF} &
  \cellcolor[HTML]{EFEFEF} &
  \cellcolor[HTML]{EFEFEF} &
  \cellcolor[HTML]{EFEFEF}Background (12000, 51\%) \\
 &
  \multirow{-2}{*}{\cellcolor[HTML]{EFEFEF}train} &
  \multirow{-2}{*}{\cellcolor[HTML]{EFEFEF}106} &
  \multirow{-2}{*}{\cellcolor[HTML]{EFEFEF}23,840} &
  \multirow{-2}{*}{\cellcolor[HTML]{EFEFEF}6.1M} &
  \cellcolor[HTML]{EFEFEF}Seizure (11840, 49\%) \\
 &
   &
   &
   &
   &
  Background (5121, 71\%) \\
\multirow{-4}{*}{TUH-Seizure} &
  \multirow{-2}{*}{test} &
  \multirow{-2}{*}{32} &
  \multirow{-2}{*}{7,152} &
  \multirow{-2}{*}{1.8M} &
  Seizure (2031, 29\%) \\ \hline
\end{tabular}
\caption{\textbf{Downstream datasets}}
\label{tab:downstream-ds}
\end{table}

\textbf{Preprocessing.} The EEG data underwent a comprehensive pre-processing pipeline to ensure signal quality and standardization across recordings. Initial frequency filtering involved a high-pass filter at 0.5 Hz and a low-pass filter at 45 Hz. The signals were then down sampled to 128 Hz to reduce computational load while preserving relevant neurophysiological information. To ensure consistency across all recordings while maximizing the number of records available for pre-training, only the 16 channels common to all records were retained. These channels were systematically arranged in the following order to preserve spatial information: FP1, FP2, F7, F3, F4, F8, T7, C3, C4, T8, P7, P3, P4, P8, O1, and O2.  The average signal was then removed from each channel to reduce the effects of noise sources common to all channels and standardize reference across channels. The EEG of each record was then divided into non-overlapping windows of 16 seconds. Multiple (1024) windows were then combined into larger data "shards", used to efficiently distribute the workload on multiple GPUs in a distributed data parallel (DDP) framework during the pre-training phase.\vspace{0.35cm} 

\subsection{Model Description}\vspace{0.25cm}

\textcolor{DarkBlue}{BioSerenity-E1} is built in two main steps: (A) the continuous EEG signal is first encoded into a learnable compact representation through an EEG Tokenizer trained to reconstruct the power spectra of EEG patches using a Vector Quantized Variational Autoencoder (VQ-VAE), whose codebook is then (B) used by a second network to learn to predict the tokens associated to both masked and unmasked portions of the input EEG.\vspace{0.20cm}

The goal of the \textbf{EEG tokenizer} is to learn a way to compress the input signal into a fixed number of meaningful features able to "summarize" its most relevant aspects in a way that they can generalize to unseen EEG data. These vectorized representations are called embeddings, because they aim to "embed" in themselves the minimal information needed to reconstruct some of the features characterizing the EEG (in this case, its power spectrum). As such, EEG tokenization is the process of projecting the continuous EEG input into a common lower-dimensional discrete subspace (the "codebook") that reduces the dimensionality of the input while preserving its more relevant inner relationships.\vspace{0.20cm}

\begin{figure}[t]
\centering
\includegraphics[width=1.0\linewidth]{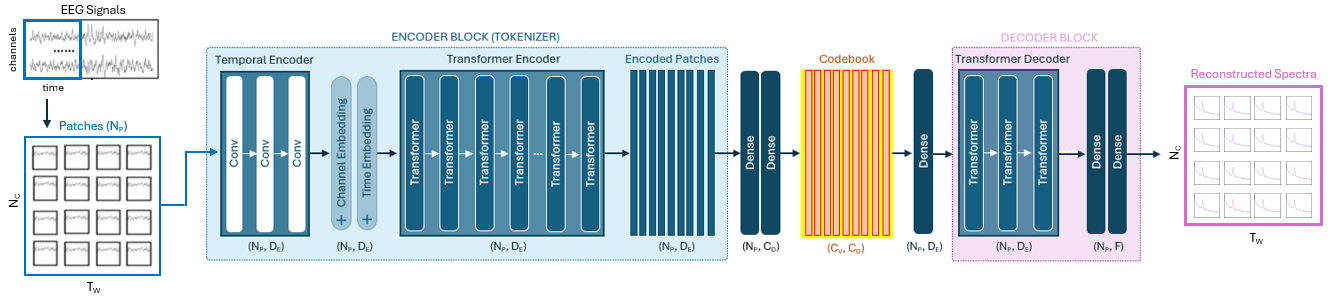}
\caption{\textbf{Tokenization and spectrum reconstruction.} EEG tokenization is based on the following VQ-VAE architecture: pre-processed EEG is segmented into windows, which are divided into patches and processed through a temporal encoder. Position and channel information is embedded using sinusoidal encoding and the resulting embeddings traverse a deep sequence of transformer blocks. The encoded vectors are then quantized by compression through dense layers and mapped to nearest codebook vectors based on cosine similarity. Finally, a shallower decoder composed of transformer blocks and dense layers reconstructs the power spectra of input patches. The trained encoder block of the VQ-VAE architecture displayed in the figure is what we refer to as Tokenizer.}
\label{fig:tkn-schema}
\end{figure}

The embeddings used to encode the EEG are learned using a transformer-based VQ-VAE, a type of neural network that combines variational autoencoders with vector quantization to learn discrete latent representations of data \cite{van2017neural, roy2018theory}. To do so, the VQ-VAE is trained to learn to encode the input data into a discrete codebook, i.e. a latent subspace of arbitrary shape that summarizes the most relevant information needed to reconstruct the power spectrum of the encoded EEG signal via a decoder block. See Figure~\ref{fig:tkn-schema}.\vspace{0.20cm}

\begin{figure}[t]
\centering
\includegraphics[width=1.0\linewidth]{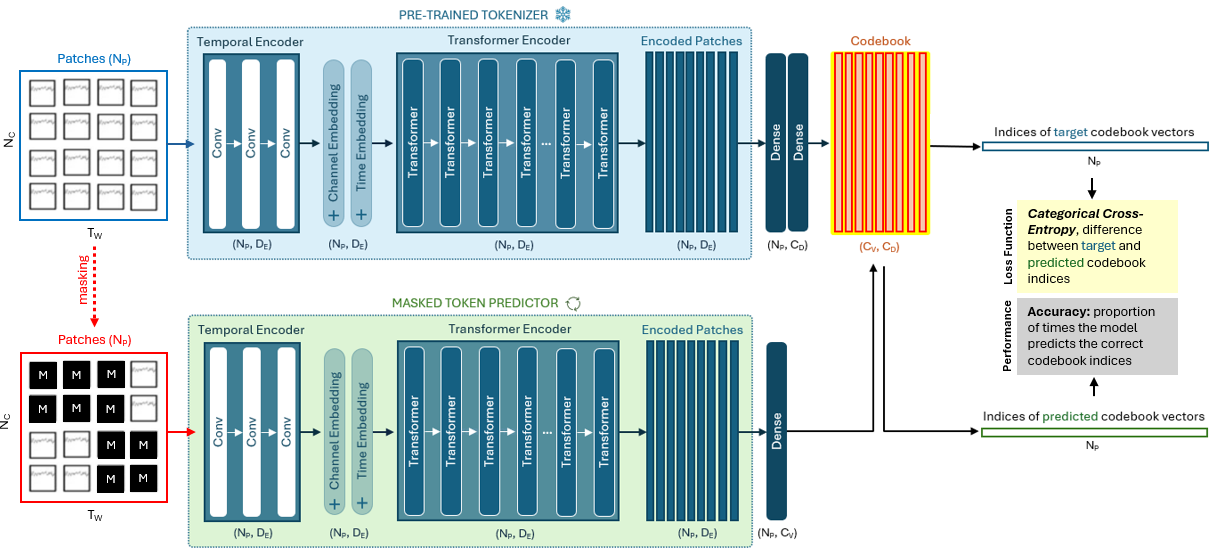}
\caption{\textbf{Masked-Token Predictor overview.} Preprocessed EEG signals are first segmented into patches and processed by the pre-trained tokenizer to obtain, for each input patch, the index corresponding to the associated latent codebook vector. These indices will be used as the correct targets to predict. A portion of the input patches are then replaced with a learnable mask and passed through a network with the same architecture as the structure of the pre-trained tokenizer but with randomly initialized weights to get the embedding vector associated to each patch. These embedding vectors storing the encoded patches are then used to obtain the indices of the predicted codebook vector associated to masked and unmasked patches. The model is trained by minimizing cross-entropy loss between predicted indices and ground truth codebook indices from the pre-trained tokenizer for both masked and unmasked patches.}
\label{fig:mtp-schema}
\end{figure}

The codebook learned by the tokenizer is then used to train a \textbf{Masked Token Predictor} (MTP) model to correctly identify the "true" codebook vectors assigned to the masked and unmasked input EEG patches (see Figure~\ref{fig:mtp-schema}).\vspace{0.35cm}   

\subsubsection{Tokenizer Architecture}\vspace{0.25cm}
The preprocessed EEG signals of each record is divided into non-overlapping, consecutive \textbf{windows} of \textbf{\textit{x}} = \{x\textit{\textsubscript{i,j }} $\in$ R\textsuperscript{TxNc} $\vert$ \textit{i} = 1, 2, …, T, \textit{j} = 1, 2, …, N\textsubscript{C}\} , where T is the number of signal samples and N\textsubscript{C} the number of EEG channels in the window. The number of samples in the window is defined as T = T\textsubscript{W} * \textit{fs}, being T\textsubscript{W} the length of the window (in seconds) and \textit{fs} the sampling frequency of the signal. Each window thus stores a total of N\textsubscript{P} = T\textsubscript{W} * N\textsubscript{C} patches, each patch being a \textit{fs}-dimensional vector storing the EEG signal corresponding to one second, one channel, \textbf{\textit{p}} = \{p\textit{\textsubscript{j,k }} $\in$ R\textsuperscript{fs} $\vert$ \textit{j} = 1, 2, …, T\textsubscript{W}, \textit{k} = 1, 2, …, N\textsubscript{C} \} . Each patch is then passed to a temporal encoder block consisting of three consecutive convolutional layers with group normalization \cite{wu2018group} and GELU activation functions \cite{hendrycks2016gaussian} to extract temporal features with output embedding dimension D\textsubscript{E} , resulting in \textbf{\textit{e}} = \{e\textit{\textsubscript{j,k }} $\in$ R\textsuperscript{D\textsubscript{E}} $\vert$ \textit{j} = 1, 2, …, T\textsubscript{W}, \textit{k} = 1, 2, …, N\textsubscript{C} \} vectors. \textit{Positional} and \textit{channel} information is then encoded into these feature tensors via an embedding layer implementing sinusoidal position encoding as proposed in \cite{vaswani2017attention}. Dropout (p=0.2) and normalization is then applied to the resulting embedding vectors, whose final shape is (N\textsubscript{P}, D\textsubscript{E}), \textbf{\textit{e}} = \{e\textit{\textsubscript{n }} $\in$ R\textsuperscript{D\textsubscript{E}} $\vert$ \textit{n} = 1, 2, …, N\textsubscript{P} \}.\vspace{0.20cm}

The embedding vectors are then fed to a sequence of 12 transformer blocks (each with 8 attention heads, hidden dimension of 1024 units and GELU activation function) which further enrich the encoding by using \textit{self-attention}, a powerful mechanism that enables the model to learn both local and global relationships in the input data \cite{vaswani2017attention}. The embedding vector \textit{e\textsubscript{n}} corresponding to each patch p\textit{\textsubscript{j,k}} (originally representing 1 second of EEG signal for 1 channel) is then mapped to \textbf{\textit{z}} = \{z\textit{\textsubscript{n}} $\in$ R\textsuperscript{D\textsubscript{E}} $\vert$ \textit{n} = 1, 2, …, N\textsubscript{P} \}. The shape of the whole \textbf{encoded sequence} being (N\textsubscript{P}, D\textsubscript{E}).\vspace{0.20cm} 

We then initialize a codebook \textbf{C} = \{v\textsubscript{i} $\vert$ \textit{i} = 1, …, C\textsubscript{V} \} $\in$ R\textsuperscript{C\textsubscript{V} x C\textsubscript{D}}, a matrix of C\textsubscript{V} discrete latent embedding vectors each defined by C\textsubscript{D} elements. These discrete vectors will be used to encode the latent representations characterizing the possible different power spectrum distributions of the EEG signal we want our model to be able to recognize and reconstruct. To do so, we first project each encoded vector in the sequence \textbf{z} from z\textsubscript{n} $\in$ R\textsuperscript{D\textsubscript{E}} to z\textsubscript{n} $\in$ R\textsuperscript{C\textsubscript{D}} (where \textit{n} = 1, …, N\textsubscript{P}) through two densely connected layers, then mapped each resulting \textit{z\textsubscript{n}} to the nearest discrete latent embedding vector v\textit{\textsubscript{i}} in the codebook based on cosine similarity, resulting in a quantized sequence with shape (N\textsubscript{P}, C\textsubscript{D}), \textbf{q = }\{q\textit{\textsubscript{n}} $\in$ R\textsuperscript{C\textsubscript{D}} $\vert$ \textit{n} = 1, 2, …, N\textsubscript{P}\}. Codebook updating is stabilized using the exponential moving average strategy and quantified by the \textit{commitment loss}, a function that prevents the encoder’s output from fluctuating too much between different codebook vectors during training \cite{van2017neural}.\vspace{0.20cm}

After quantization, the set of discretized vector embeddings \textbf{\textit{q}} corresponding to each patch in the input sequence are fed to the decoder, whose architecture mirrors the encoder but has reduced depth (3 transformer layers vs. 12 in the encoder) and a final prediction head composed by 2 sequential fully connected layers that projects the output of the transformer block for the whole sequence to reconstruct the power spectra of all patches of the input sequence with shape (N\textsubscript{P}, F), being F the number of spectral frequencies, \textbf{\textit{o}} = \{\textit{o\textsubscript{n}} $\in$ R\textsuperscript{F} $\vert$ \textit{n} = 1, 2, …, N\textsubscript{P}\}. The output \textbf{\textit{o}} of the decoder will then be used to calculate the \textit{reconstruction loss} against the target power spectrum of each patch, \textbf{\textit{s}} = \{\textit{s\textsubscript{n}} $\in$ R\textsuperscript{F} $\vert$ \textit{n} = 1, 2, …, N\textsubscript{P}\}, quantified using mean squared error. The power spectrum \textit{s\textsubscript{n}} of each patch was estimated using Discrete Prolate Spheroidal Sequences (DPSS) multitaper windowing \cite{slepian1961prolate, percival1993spectral}. The trained encoder block of the described VQ-VAE is what we refer to as “EEG tokenizer”, as its output are the embedding vectors (the “tokens”) encoding the EEG input patches.\vspace{0.35cm}

\subsubsection{Masked Token Predictor Architecture}\vspace{0.25cm}
The pre-trained EEG Tokenizer described above is used to obtain the "ground-truth" that the masked token predictor model will be trained to predict using masked self-supervised learning. The preprocessed EEG signals are first segmented into patches and processed by the pre-trained tokenizer to obtain, for each input patch, the index corresponding to the associated \textit{target} vector in the codebook. The MTP is based upon the encoder architecture used in the VQ-VAE-based tokenizer described above. EEG signal is preprocessed and then segmented into a sequence of patches, which are later transformed via a temporal encoder layer to extract temporal feature tensors for each patch, \textbf{\textit{e}} = \{e\textit{\textsubscript{j,k}} $\vert$ \textit{j} = 1, 2, …, T\textsubscript{W}, \textit{k} = 1, 2, …, N\textsubscript{C}\}, as described above. A given portion \textit{r} of these embeddings were randomly substituted via a learnable mask e\textsubscript{M} $\in$ R\textsuperscript{D\textsubscript{E}} token initialized with small random values drawn from a normal distribution, and the resulting tensor (storing both masked and non-masked patch embedding vectors) is then enriched via position and channel encoding using the same embedding layer architecture used by the tokenizer. The resulting embedding vectors are passed to a sequence of 12 transformer layers using GELU activation functions and 16 attention heads, resulting in the output sequence \textbf{\textit{z}} = \{z\textit{\textsubscript{n}} $\in$ R\textsuperscript{D\textsubscript{E}} $\vert$ \textit{n} = 1, 2, …, N\textsubscript{P}\}. The output sequence is processed by a multi-layer perceptron head that first projects the transformer output for each patch into a C\textsubscript{V}-dimensional vector of logits, \textbf{\textit{h}} = \{h\textit{\textsubscript{n }} $\in$ R\textsuperscript{C\textsubscript{V}} $\vert$ \textit{n} = 1, 2, …, N\textsubscript{P}\}, which are then used to predict the index of the codebook vector associated to the largest logit for each patch. The whole MTP model is trained by minimizing the \textit{cross-entropy loss function} between the true indices associated to each patch in the sequence (obtained from the codebook of the pre-trained tokenizer model) and the predicted ones for both the masked and the unmasked patches.\vspace{0.35cm}

\subsection{Model Training}\vspace{0.25cm}
\textcolor{DarkBlue}{BioSerenity-E1} undergoes a two-phase training process. Initially, it is pre-trained through self-supervised representation learning, which involves Tokenization and Partial Masking of the input EEG signals. Following this pre-training phase, the model is fine-tuned on specific supervised learning objectives tailored to distinct clinical applications, including normal versus abnormal EEG classification, disease prediction, and seizure detection.\vspace{0.25cm}

\subsubsection{Tokenizer Pre-Training}\vspace{0.25cm}
The tokenizer model was trained for 100 epochs using AdamW optimizer \cite{adamwoptimizer} with a cosine learning rate scheduler \cite{cosinelr2016} using batch size of 128. The tokenizer optimizes a loss function defined as the sum of both the spectrum reconstruction loss and the commitment loss from vector quantization. Examples of the original vs. reconstructed power spectra obtained after tokenizer pre-training are in Figure~\ref{fig:tkn-spectra} in the Appendix. To assess the extent to which the tokenizer utilized its representational capacity during pre-training, we also quantified the codebook usage percentage—defined as the proportion of codebook vectors actively used—and the normalized codebook entropy, which measures how evenly these utilized vectors are distributed. The tokenizer's pre-training metrics over epochs can be seen in Figure~\ref{fig:tkn-pretrain-curves} in the Appendix. The transformer-based VQ-VAE used to build \textcolor{DarkBlue}{BioSerenity-E1} has 12.6M learnable parameters and was pre-trained on 4000 EEG hours (Table~\ref{tab:pretrain-ds}).  The hyperparameters values of the tokenizer model are shown in Table~\ref{tab:hps-tkn} in the Appendix.

\subsubsection{Masked Token Predictor Pre-Training}\vspace{0.25cm}
The masked-token prediction model was trained for 30 epochs using AdamW optimizer with a cosine learning rate scheduler using batch size of 128. The MTP is trained to minimize the cross-entropy loss between the predicted codebook indices assigned to masked and unmasked input patches and the true indices assigned to the unmasked input patches by the pretrained tokenizer's encoder. As additional MTP training performance metric we also quantified the accuracy over epochs between the predicted and the true codebook indices. MTP pre-training metrics over epochs can be seen in Figure~\ref{fig:mtp-pretrain-curves} in the Appendix. The masked-token-prediction model has 11.7M learnable parameters and was pre-trained on 4000 EEG hours (Table~\ref{tab:pretrain-ds}). MTP’s hyperparameters values are shown in Table~\ref{tab:hps-mtp} in the Appendix. The pre-trained masked token predictor model described so far is what we refer to as \textcolor{DarkBlue}{BioSerenity-E1}.

\subsubsection{Fine-Tuning}
\begin{figure}[h]
\centering
\includegraphics[width=1.0\linewidth]{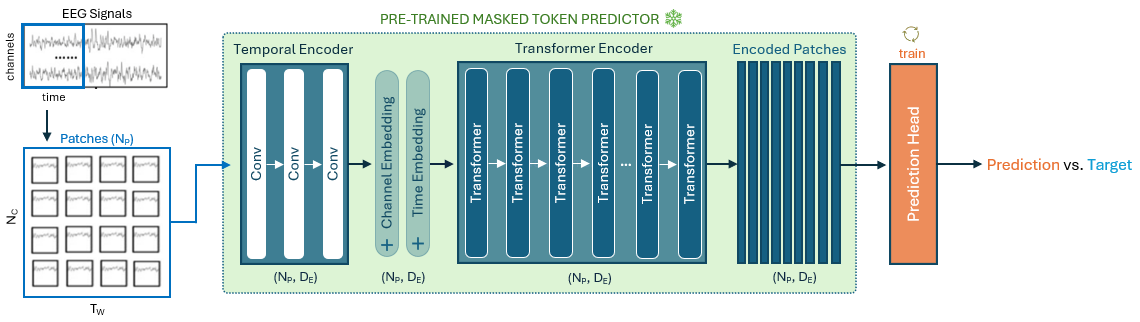}
\caption{\textbf{Fine-tuning overview.} \textcolor{DarkBlue}{BioSerenity-E1} (i.e. the pre-trained Masked-Token Predictor) serves as base model for a trainable prediction head that is trained on the downstream task of interest. To accelerate fine-tuning, we froze all base model's weights, keeping only the prediction head trainable.}
\label{fig:finetune-schema}
\end{figure}

\vspace{0.25cm}
\textcolor{DarkBlue}{BioSerenity-E1} was used as base model on multiple downstream datasets to evaluate generalizability of the learned representations. To do so, we froze \textcolor{DarkBlue}{BioSerenity-E1}'s model weights and added a prediction head composed by three convolutional layers and an average pooling layer that was trained via supervised learning to predict the unique label associated to each input window of the corresponding downstream training dataset (Figure~\ref{fig:finetune-schema}). Fine-tuning was performed using early stopping (patience = 5 epochs), batch size of 128, Adam optimizer and constant learning rate (LR\textsubscript{head} = 1e-04). We used binary cross-entropy as loss function for \textit{Neurophy-Abnormal}, \textit{TUH-Abnormal} and \textit{TUH-Seizure} and cross-entropy for \textit{Neurophy-Multiclass.} To assess the variability of results due to randomness in fine-tuning, we run three finetuning jobs for each model and downstream task.

\subsubsection{Baseline Models}\vspace{0.25cm}
We evaluated \textcolor{DarkBlue}{BioSerenity-E1} against three baseline models and then compared its performance with several state-of-the-art published results. \textbf{MTP-Virgin}, corresponding to a 12M parameters model having the same architecture as \textcolor{DarkBlue}{BioSerenity-E1} but that is trained from-scratch on the training data of each downstream task, used to show the contribution of our pre-training strategy. \textbf{EEGWNet,} a sophisticated network tailored to extract and focus on hierarchical temporal features using parallel 1D Convolutional blocks and residual layers that has been inspired by GWNET \cite{dhankhargwnet}, characterized by approximately 180K learnable parameters. \textbf{FC3Net, }a compact network composed of 3 fully connected layers with 64, 32 and 32 neurons respectively, with two dropout gates (0.2 and 0.5) and GELU activation functions, with approximately 12K learnable parameters. \textbf{DeepSOZ} is a network that combines a transformer encoder with an LSTM block, as described by \cite{shama2023deepsoz}. Another model, the Temporal Graph Convolutional Network (\textbf{TGCN} \cite{covert2019temporal}), was developed for handling temporal data in graphs. \textbf{CNN-BLSTM} is a model that was proposed for long-term seizure monitoring \cite{craley2021automated}, while \textbf{TSD} is a transformer-based model specifically designed for seizure detection, developed by \cite{ma2023tsd}. \textbf{ConvLSTM} consist of multiple ConvLSTM blocks followed by fully connected layers \cite{yang2021continental}. \textbf{Dist-DCRNN} is a pre-trained graph neural network, as detailed by \cite{tang2021automated}. \textbf{EEGNet}, developed by \cite{lawhern2018eegnet}, is a compact CNN tailored for BCI applications. The Temporal Convolutional Network (\textbf{TCN}) uses dilated convolutional neural networks, as introduced by \cite{bai2018empirical}. \textbf{EEG-GNN}, developed by \cite{tang2021self}, utilizes graph neural networks to capture spatiotemporal dependencies in EEG data. \textbf{GraphS4mer},  by \cite{tang2023modeling}, incorporates structured state space models for multivariate biosignals. \textbf{BrainBERT}, developed by \cite{wang2023brainbert}, employs neural signal processing techniques to produce superresolution time-frequency representations and is pre-trained with mask reconstruction loss. \textbf{EEGFormer,} introduced by \cite{chen2024eegformer}, is a family of pre-trained EEG foundation models based on a transformer encoder-decoder architecture. It includes variants like EEGFormer-s, EEGFormer-b, and EEGFormer-l, each with different encoder layers and codebook vectors. \textbf{CBRaMod} is another EEG foundation model, pre-trained on 27,000 hours of EEG data and featuring criss-cross attention with asymmetric conditional positional encoding \cite{wang2024cbramod}. \textbf{LaBraM}, developed by \cite{labram2024}, is an EEG foundation model pre-trained on 2,500 hours of EEG data, available in three sizes: Base (5.8M), Large (46M), and Huge (369M). \textbf{BIOT}, introduced by \cite{biot2023}, is a generic biosignal learning model that tokenizes diverse biosignals into unified "sentences." This model was pre-trained on 58,020 hours of biosignals, including EEG data. Five other supervised methods are also utilized as baselines: \textbf{SPaRCNet} \cite{jing2023development}, \textbf{ContraWR} \cite{yang2021self}, \textbf{CNN-Transformer} \cite{peh2022transformer}, \textbf{FFCL} \cite{li2022motor}, and \textbf{ST-Transformer} \cite{song2021transformer}.

\subsubsection{Evaluation Metrics}\vspace{0.25cm}
We adopted a number of metrics to assess performance on both binary and multi-class classification downstream tasks. \textbf{Sensitivity}, also known as the True Positive Rate, evaluates a model's ability to correctly identify positive cases. It measures the proportion of actual positive cases that were correctly identified. \textbf{Specificity}, also referred to as True Negative Rate, measures the model's ability to correctly identify negative cases, calculated as the ratio of true negatives to all negative outcomes. This metric is particularly important when there's a high cost associated with false positives. \textbf{Accuracy} (used during pre-training of the masked-token predictor) is the ratio of correctly predicted instances to the total number of instances. \textbf{Balanced Accuracy} is the arithmetic mean of sensitivity and specificity, particularly useful on imbalanced datasets. The Area Under the Receiver Operating Characteristic curve (\textbf{AUROC}) measures the overall model's discriminatory ability in terms of True Positive Rate and False Positive Rate assessed on a range of classification thresholds. Values range from 0.5 (random guess) to 1.0 (perfect classification). \textbf{AUPRC} is a performance metric calculating the area under the Precision-Recall (PR) curve obtained over a range of classification thresholds. It summarizes the trade-off between Precision (accuracy of positive predictions) and Recall (ability to find all positive cases). \textbf{F1 score} is the harmonic mean of Precision and Recall; in the resent paper, we always used the \textit{weighted F1 score}, that accounts for class imbalance by weighting each class's F1 score based on its frequency in the dataset. Together with AUPRC, the weighted F1 score provides a robust measure in multi-class classification settings, where different classes may have different prevalence. \vspace{0.25cm}

\subsubsection{Implementation Details}\vspace{0.25cm}
Model building and training were implemented in PyTorch \cite{paszke2019pytorch} using BF16 floating-point precision. The tokenizer is characterized by 12.6M learnable parameters and was trained on 4000 hours of EEG data using distributed data parallel (DDP) on 16 NVIDIA A10G GPUs for 100 epochs with a batch size of 128. Tokenization training took 6.7 hours. The masked-token predictor model has 11.7M learnable parameters and was trained on 4000 hours of EEG data using distributed data parallel (DDP) on 16 NVIDIA A10G GPUs for 30 epochs with a batch size of 128. Masked-token prediction training took 3.5 hours. Fine-tuning was performed on a single NVIDIA A10G GPU the batch size was set to 32, with the total training time per epoch ranging from 5 to 30 minutes, depending on the task.\vspace{0.35cm}

\clearpage 
\section{Results}

\subsection{Seizure Detection}
\begin{figure}[h]
\centering
\includegraphics[width=1.0\linewidth]{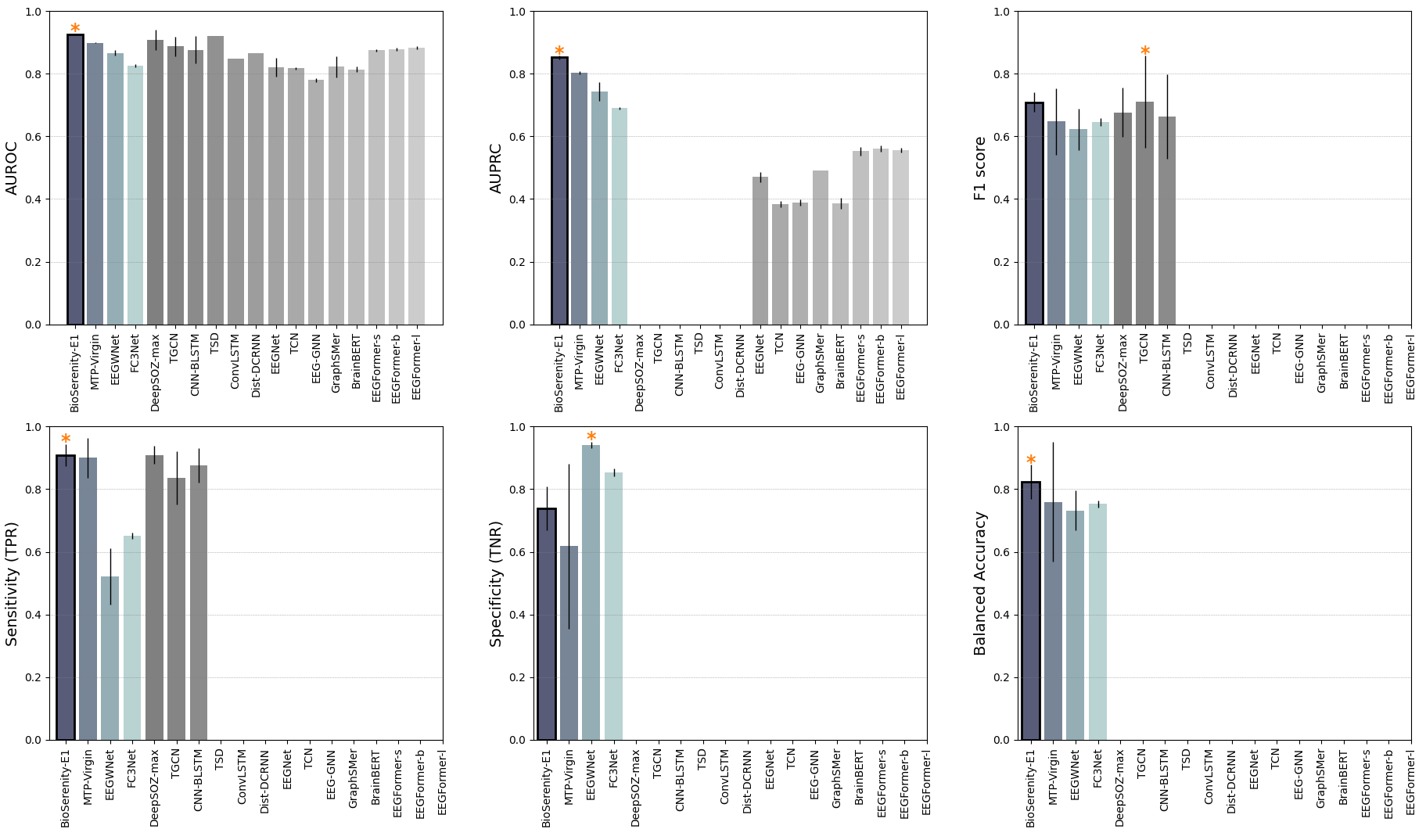}
\caption{\textbf{Seizure Detection on TUH-Seizure.} Performance comparison of \textcolor{DarkBlue}{BioSerenity-E1} and baseline models on the seizure detection task using the TUH-Seizure dataset. \textcolor{DarkBlue}{BioSerenity-E1} demonstrates superior performance across most metrics, achieving the highest AUROC, AUPRC, Sensitivity, and Balanced Accuracy (indicated by stars). Error bars represent standard deviations over multiple runs. Models in shades of blue were run as baseline models to evaluate our model, whereas results of those in shades of grey represent the state-of-the-art obtained from the literature for binary seizure detection (see section “Baseline Models”).}
\label{fig:results-tusz}
\end{figure}

\vspace{0.25cm}
The performance of \textcolor{DarkBlue}{BioSerenity-E1} was evaluated against several baseline models on the seizure detection task using the TUH-Seizure dataset. Results indicate that \textcolor{DarkBlue}{BioSerenity-E1} consistently outperformed other models across key metrics (Figure \ref{fig:results-tusz}). Specifically, it achieved the highest AUROC, AUPRC, Sensitivity (TPR), and Balanced Accuracy, as highlighted in the plots (stars indicate top performance). The model also demonstrated competitive F1-scores and Specificity (TNR), maintaining a robust balance between true positive and true negative rates. These results underscore the effectiveness of \textcolor{DarkBlue}{BioSerenity-E1} as a foundation model for EEG-based seizure detection tasks. Error bars in the plots reflect variability across multiple runs, further validating the model's reliability.\vspace{0.35cm}

\subsection{Normal vs. Abnormal EEG Classification}
\begin{figure}[th]
\centering
\includegraphics[width=1.0\linewidth]{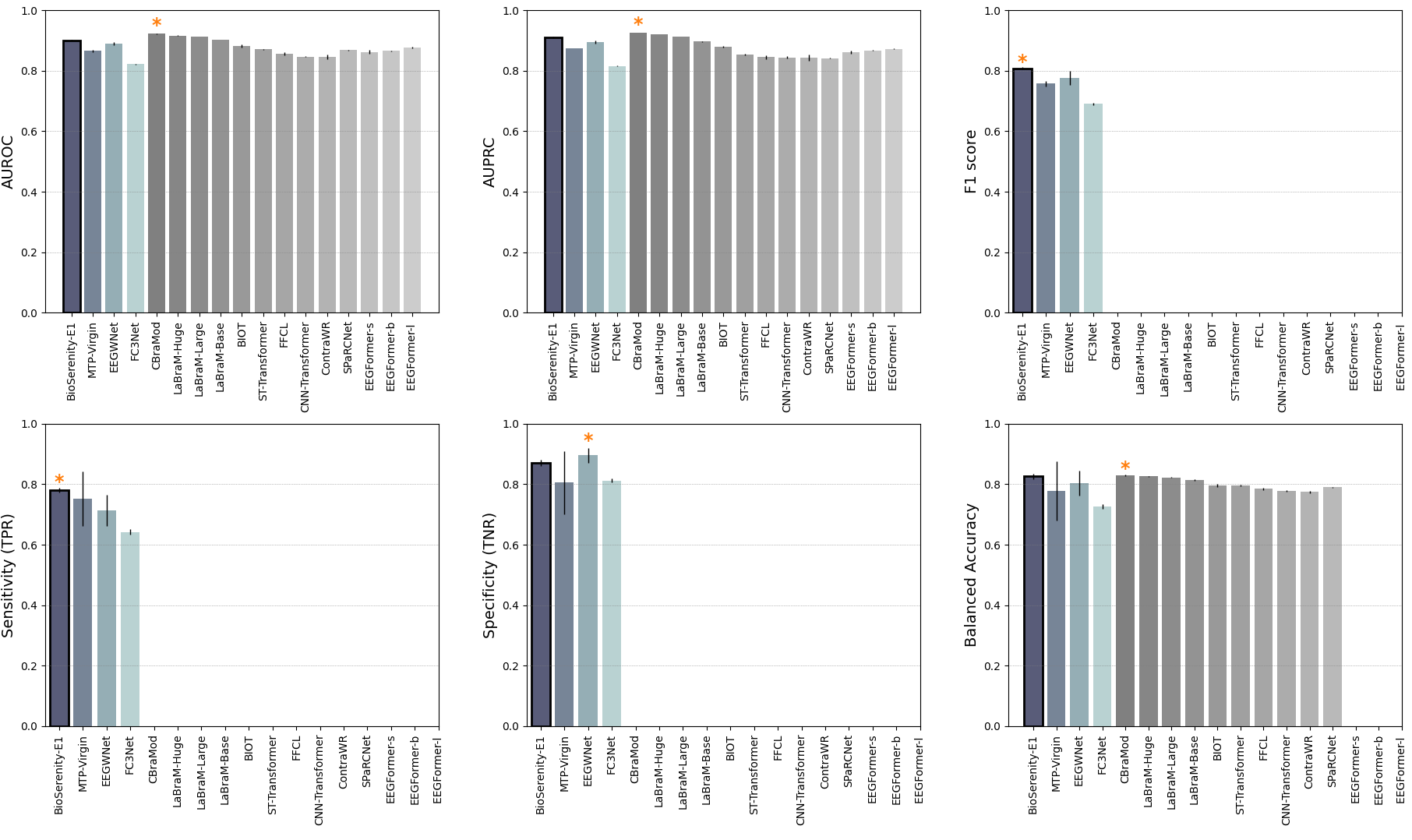}
\caption{\textbf{Normal vs. Abnormal Classification on TUH-Abnormal.} Performance comparison of \textcolor{DarkBlue}{BioSerenity-E1} against baseline models on the TUH-Abnormal EEG dataset for the normal-vs-abnormal classification task. \textcolor{DarkBlue}{BioSerenity-E1} ranks in the top quartile of state-of-the-art across all metrics, achieving the highest weighted F1 scores and Sensitivity (TPR), as indicated by the orange star. Models in shades of blue were run as baseline models to evaluate our model, whereas results of those in shades of grey represent the state-of-the-art obtained from the literature for binary seizure detection (see section “Baseline Models”).}
\label{fig:results-tuab}
\end{figure}

\begin{figure}[h]
\centering
\includegraphics[width=1.0\linewidth]{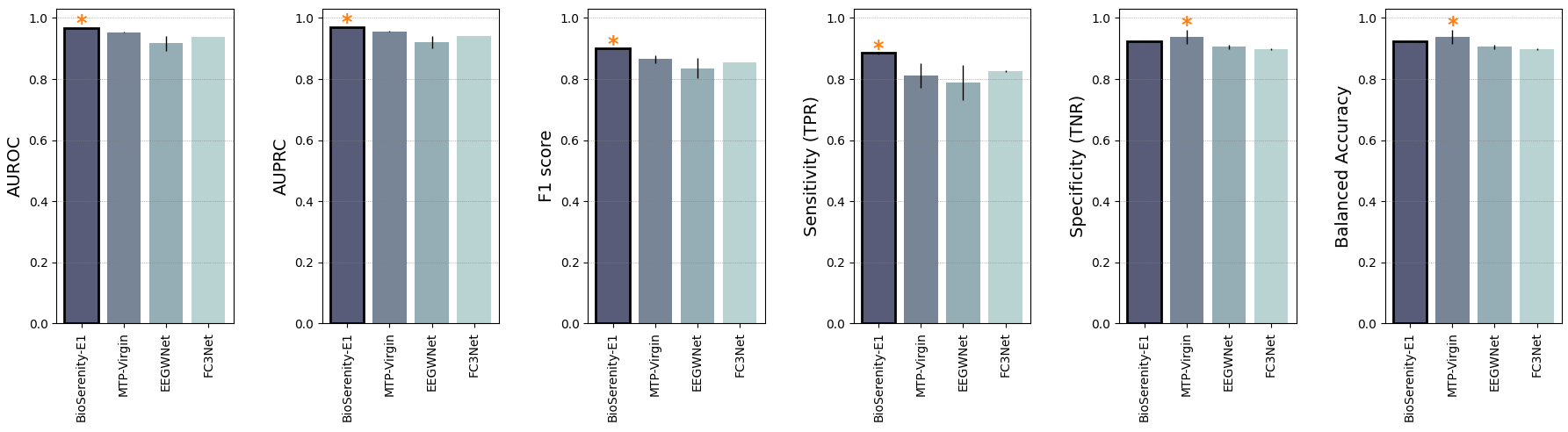}
\caption{\textbf{Normal vs. Abnormal Classification on Neurophy-Abnormal.} Performance comparison of \textcolor{DarkBlue}{BioSerenity-E1} against three baseline models on the normal-vs-abnormal EEG classification task using the proprietary Neurophy-Abnormal dataset. \textcolor{DarkBlue}{BioSerenity-E1} outperforms all baseline models across most metrics, ranking second on Specificity (TNR) and Balanced Accuracy (orange asterisks mark best result). State-of-the-art performances from the literature are not available as “Neurophy-Abnormal” is a proprietary dataset (See “Downstream Datasets”).}
\label{fig:results-neurophy-abnormal}
\end{figure}

\vspace{0.25cm}
We evaluated \textcolor{DarkBlue}{BioSerenity-E1} performance in correctly distinguishing normal against clinically abnormal EEG using two datasets: TUH-Abnormal and Neurophy-Abnormal. On TUH-Abnormal (Figure \ref{fig:results-tuab}), \textcolor{DarkBlue}{BioSerenity-E1} consistently ranks in top-tier across all metrics, demonstrating competitive performance. \textcolor{DarkBlue}{BioSerenity-E1} excels in sensitivity (TPR) and weighted F1 score, highlighting its ability to correctly identify abnormal EEGs and maintain a strong balance between precision and recall. CBraMod achieves better results than \textcolor{DarkBlue}{BioSerenity-E1} in AUROC, AUPRC, and balanced accuracy. Similarly, EEGWNet outperforms \textcolor{DarkBlue}{BioSerenity-E1} in specificity (TNR). It should be noted that the differences in TUH-Abnormal performance across top-tier models are, however, suggesting that the state-of-the-art is possibly touching roof on this specific dataset. \textcolor{DarkBlue}{BioSerenity-E1} exhibited also excellent classification performances on the proprietary Neurophy-Abnormal dataset (Figure \ref{fig:results-neurophy-abnormal}), scoring second to MTP-Virgin only on Specificity (TNR) and Balanced Accuracy, showing however a small percentage difference and displaying higher stability, as showed by decreased inter-trial variability compared to MTP-Virgin.\vspace{0.35cm}

\subsection{Multiclass EEG Classification}
\begin{figure}[h]
\centering
\includegraphics[width=1.0\linewidth]{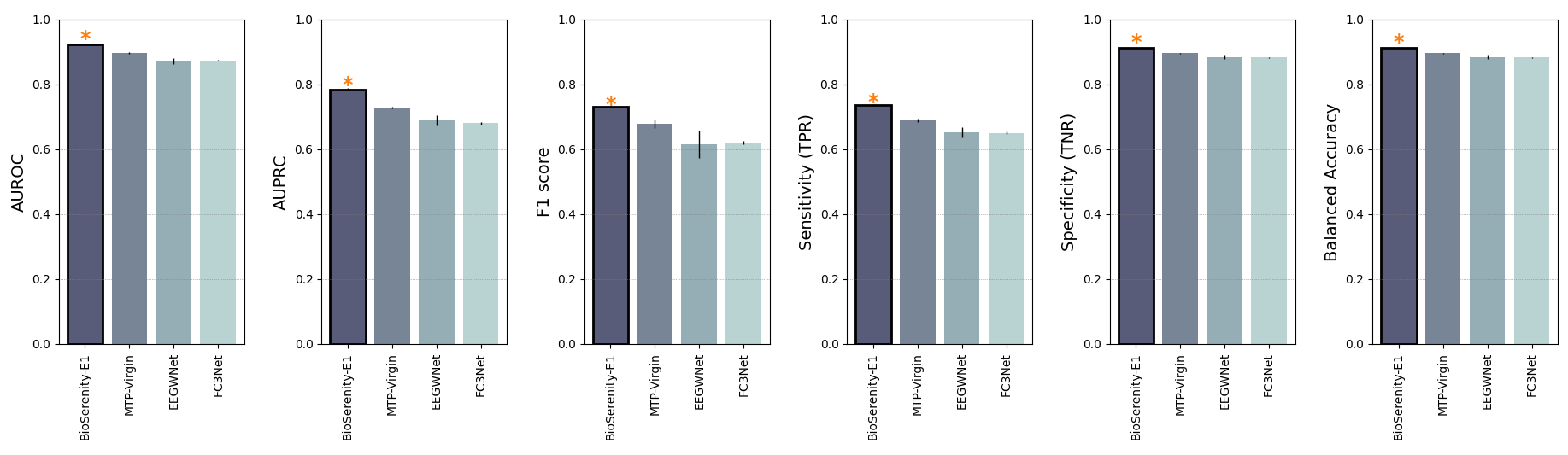}
\caption{\textbf{Abnormality Classification on Neurophy-Multiclass.} Performance comparison of \textcolor{DarkBlue}{BioSerenity-E1} against three baseline models on the multi-class EEG classification task using the proprietary Neurophy-Multiclass dataset. Each window in this dataset is assigned to a unique label, that the models are trained to predict. The target classes are: “normal”, “status epilepticus”, “lesion” and “encephalopathy”. \textcolor{DarkBlue}{BioSerenity-E1} consistently outperforms all baseline models across all metrics (orange asterisks mark best result). State-of-the-art performances from the literature are not available as “Neurophy-Multiclass” is a proprietary dataset (See “Downstream Datasets”).}
\label{fig:results-neurophy-multi}
\end{figure}

\vspace{0.25cm}
The performance of \textcolor{DarkBlue}{BioSerenity-E1} was also evaluated on the multiclass EEG classification task using the proprietary Neurophy-Multiclass dataset and compared against three internal baseline models: MTP-Virgin, EEGWNet, and FC3Net. As shown in Figure~\ref{fig:results-neurophy-multi}, \textcolor{DarkBlue}{BioSerenity-E1} demonstrated superior performance across all evaluated metrics: it achieved the highest AUROC, indicating its robust ability to distinguish between classes. The largest improvements compared to the other baseline models are seen in AUPRC and F1 score, highlighting \textcolor{DarkBlue}{BioSerenity-E1} effectiveness in handling imbalance in the training data (see Table~\ref{tab:downstream-ds}) and optimizing precision-recall trade-offs, two important aspects to control when designing algorithms to support clinical applications.

\subsection{Performance in Low-Data Regimes}
\begin{figure}[t!]
\centering
\includegraphics[width=1.0\linewidth]{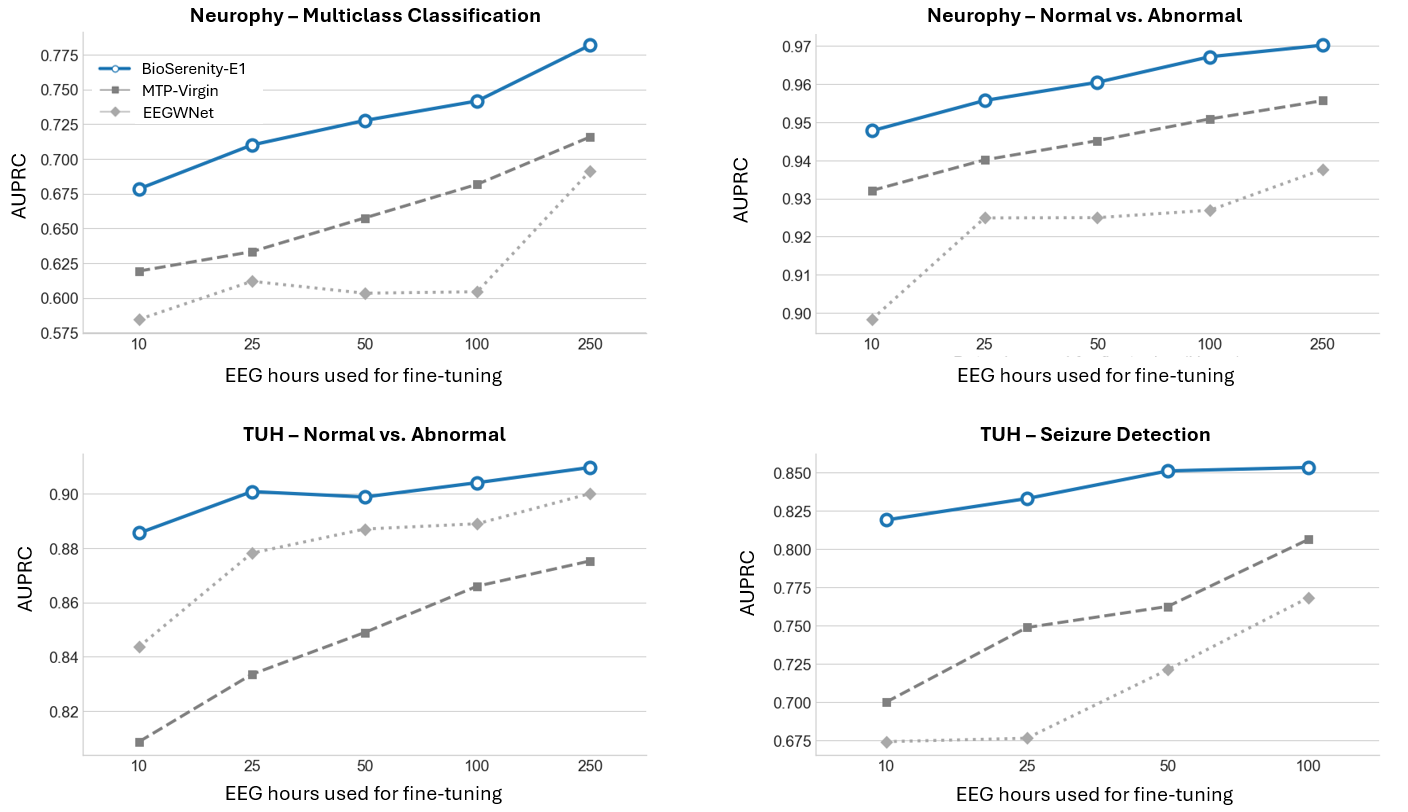}
\caption{\textbf{Performance comparison in low-data regimes across downstream tasks.} The plot shows the performance improvements of \textcolor{DarkBlue}{BioSerenity-E1} over MTP-Virgin and EEGWNet when trained on small fractions of the available training data across the four downstream datasets.}
\label{fig:low-data}
\end{figure}

\vspace{0.25cm}
One of the most limiting factors in the development of clinical AI algorithms is data scarcity, either because obtaining enough data of good quality and annotations can be expensive and time-consuming, or because the prevalence of the condition under study is low (e.g. in the case of rare diseases) and therefore the available data volumetry on which to train is intrinsically small. To assess the utility of \textcolor{DarkBlue}{BioSerenity-E1} in low-data regimes, we fine-tuned it using only fractions of the available training data and quantified its performance against the full test set on the downstream tasks (see Figure~\ref{fig:low-data}). We observed the positive impact of the pre-trained \textcolor{DarkBlue}{BioSerenity-E1} in low-data regimes, against a structurally identical but pristine network architecture (MTP-Virgin) as well as against the sophisticated EEGWNet.  As an example, the relative increase of using \textcolor{DarkBlue}{BioSerenity-E1} against the second-best baseline model (i.e. MTP-Virgin) using as few as 10 hours of training data is +9\% on Neurophy-Multiclass, +2\% on Neurophy-Abnormal, 5\% on TUH-Abnormal, and +17\% on TUH-Seizure.\vspace{0.35cm}

\section{Conclusions}
Here we present \textcolor{DarkBlue}{BioSerenity-E1}, a self-supervised EEG model that leverages EEG Tokenization and Masked Token Prediction. Pre-trained on 4,000 hours of EEG data obtained from clinical settings, \textcolor{DarkBlue}{BioSerenity-E1} is designed to serve as a foundation model to accelerate development and improve performance of EEG-based medical applications. The model was evaluated on three relevant use cases (seizure detection, EEG abnormality detection and general disease classification) using two open-source and two proprietary datasets. Performance was compared against established model baselines and state-of-the-art results from the literature using a comprehensive set of metrics. Despite being pre-trained on moderate data volumetry, results illustrate that \textcolor{DarkBlue}{BioSerenity-E1} exhibits performances that are either ranking in the top-tier of the state of the art (see Normal vs. Abnormal EEG classification on the TUH-Abnormal dataset) or beating it (see Seizure Detection on TUH-Seizure). Future work will assess its performance on other clinically relevant downstream tasks and datasets. \textcolor{DarkBlue}{BioSerenity-E1} was developed for clinical applications utilizing the standard 10-20 EEG system. With this objective we selected the largest possible set of channels from available databases to maximize the total hours of EEG data available for pre-training. The results presented in the current study are obtained using models trained on the resulting channels. Future work will address this limitation to make the model flexible in terms of input channels. Another limit of the current model is the reliance on a relatively small and possibly too homogeneous dataset for pretraining. This constraint may result in an underutilized VQ-VAE codebook, as the model is not exposed to a diverse enough set of EEG patterns and conditions during training. In fact, only less than 20\% of the codebook vectors tend to be actively used in the tokenizer, even though the used vectors do have a relatively even distribution ($\sim$80\% of the maximum possible entropy). Consequently, the model might not fully leverage the representational capacity of the codebook to capture complex and varied EEG features, potentially limiting its ability to generalize effectively across different populations and experimental settings. Future work should focus on expanding the dataset to include more diverse EEG recordings, which could enhance the model's robustness and versatility. In addition, overall codebook usage could be improved through entropy regularization techniques that encourage more balanced codebook utilization while maintaining reconstruction quality \cite{volkov2022homology, baykal2024edvae}.\vspace{0.35cm} 

\section{Author Contributions}
R.G.B. and M.R. contributed equally to designing the project, M.R. implemented data parallelization, tokenizer pre-training, experiments and fine-tuning pipelines, R.G.B. implemented data extraction, masked token prediction pre-training and wrote the manuscript, M.R. and U.G. reviewed the manuscript. All authors reviewed and approved the final manuscript.\vspace{0.35cm}

\clearpage
\bibliographystyle{apalike}
\bibliography{bibliography.bib}

\begin{thebibliography}{}

\bibitem[Abibullaev et~al., 2023]{abibullaev2023deep}
Abibullaev, B., Keutayeva, A., and Zollanvari, A. (2023).
\newblock Deep learning in eeg-based bcis: A comprehensive review of transformer models, advantages, challenges, and applications.
\newblock {\em IEEE Access}, 11:127271--127301.

\bibitem[Alhagry et~al., 2017]{alhagry2017emotion}
Alhagry, S., Fahmy, A.~A., and El-Khoribi, R.~A. (2017).
\newblock Emotion recognition based on eeg using lstm recurrent neural network.
\newblock {\em International Journal of Advanced Computer Science and Applications}, 8(10).

\bibitem[Bai et~al., 2018]{bai2018empirical}
Bai, S., Kolter, J.~Z., and Koltun, V. (2018).
\newblock An empirical evaluation of generic convolutional and recurrent networks for sequence modeling.
\newblock {\em arXiv preprint arXiv:1803.01271}.

\bibitem[Baykal et~al., 2024]{baykal2024edvae}
Baykal, G., Kandemir, M., and Unal, G. (2024).
\newblock Edvae: Mitigating codebook collapse with evidential discrete variational autoencoders.
\newblock {\em Pattern Recognition}, 156:110792.

\bibitem[Behzad and Behzad, 2021]{behzad2021role}
Behzad, R. and Behzad, A. (2021).
\newblock The role of eeg in the diagnosis and management of patients with sleep disorders.
\newblock {\em Journal of Behavioral and Brain Science}, 11(10):257--266.

\bibitem[Bera, 2021]{bera2021}
Bera, T.~K. (2021).
\newblock A review on the medical applications of electroencephalography (eeg).
\newblock In {\em 2021 Seventh International conference on Bio Signals, Images, and Instrumentation (ICBSII)}, pages 1--6.

\bibitem[Bettinardi, 2016]{bettinardi2016spontaneous}
Bettinardi, R.~G. (2016).
\newblock Spontaneous brain activity: how dynamics and topology shape the emergent correlation structure.
\newblock {\em PhD Thesis}.

\bibitem[Blanco et~al., 1995]{blanco1995stationarity}
Blanco, S., Garcia, H., Quiroga, R.~Q., Romanelli, L., and Rosso, O. (1995).
\newblock Stationarity of the eeg series.
\newblock {\em IEEE Engineering in medicine and biology Magazine}, 14(4):395--399.

\bibitem[Blinowska and Durka, 2006]{blinowska2006}
Blinowska, K. and Durka, P. (2006).
\newblock Electroencephalography (eeg).
\newblock {\em Wiley encyclopedia of biomedical engineering}, 10:9780471740360.

\bibitem[Cabral et~al., 2014]{cabral2014exploring}
Cabral, J., Kringelbach, M.~L., and Deco, G. (2014).
\newblock Exploring the network dynamics underlying brain activity during rest.
\newblock {\em Progress in Neurobiology}, 114:102--131.

\bibitem[Cai et~al., 2023]{cai2023mbrain}
Cai, D., Chen, J., Yang, Y., Liu, T., and Li, Y. (2023).
\newblock Mbrain: A multi-channel self-supervised learning framework for brain signals.
\newblock In {\em Proceedings of the 29th ACM SIGKDD Conference on Knowledge Discovery and Data Mining}, pages 130--141.

\bibitem[Chen et~al., 2024]{chen2024eegformer}
Chen, Y., Ren, K., Song, K., Wang, Y., Wang, Y., Li, D., and Qiu, L. (2024).
\newblock Eegformer: Towards transferable and interpretable large-scale eeg foundation model.
\newblock {\em arXiv preprint arXiv:2401.10278}.

\bibitem[Chien et~al., 2022]{chien2022maeeg}
Chien, H.-Y.~S., Goh, H., Sandino, C.~M., and Cheng, J.~Y. (2022).
\newblock Maeeg: Masked auto-encoder for eeg representation learning.
\newblock {\em arXiv preprint arXiv:2211.02625}.

\bibitem[Coburn et~al., 2006]{coburn2006value}
Coburn, K.~L., Lauterbach, E.~C., Boutros, N.~N., Black, K.~J., Arciniegas, D.~B., and Coffey, C.~E. (2006).
\newblock The value of quantitative electroencephalography in clinical psychiatry: a report by the committee on research of the american neuropsychiatric association.
\newblock {\em The Journal of neuropsychiatry and clinical neurosciences}, 18(4):460--500.

\bibitem[Covert et~al., 2019]{covert2019temporal}
Covert, I.~C., Krishnan, B., Najm, I., Zhan, J., Shore, M., Hixson, J., and Po, M.~J. (2019).
\newblock Temporal graph convolutional networks for automatic seizure detection.
\newblock In {\em Machine learning for healthcare conference}, pages 160--180. PMLR.

\bibitem[Craik et~al., 2019]{craik2019deep}
Craik, A., He, Y., and Contreras-Vidal, J.~L. (2019).
\newblock Deep learning for electroencephalogram (eeg) classification tasks: a review.
\newblock {\em Journal of neural engineering}, 16(3):031001.

\bibitem[Craley et~al., 2021]{craley2021automated}
Craley, J., Johnson, E., Jouny, C., and Venkataraman, A. (2021).
\newblock Automated inter-patient seizure detection using multichannel convolutional and recurrent neural networks.
\newblock {\em Biomedical signal processing and control}, 64:102360.

\bibitem[Cruzat et~al., 2023]{cruzat2023temporal}
Cruzat, J., Herzog, R., Prado, P., Sanz-Perl, Y., Gonzalez-Gomez, R., Moguilner, S., Kringelbach, M.~L., Deco, G., Tagliazucchi, E., and Iba{\~n}ez, A. (2023).
\newblock Temporal irreversibility of large-scale brain dynamics in alzheimer’s disease.
\newblock {\em Journal of Neuroscience}, 43(9):1643--1656.

\bibitem[Cui et~al., 2024]{cui2024neuro}
Cui, W., Jeong, W., Th{\"o}lke, P., Medani, T., Jerbi, K., Joshi, A.~A., and Leahy, R.~M. (2024).
\newblock Neuro-gpt: Towards a foundation model for eeg.
\newblock In {\em 2024 IEEE International Symposium on Biomedical Imaging (ISBI)}, pages 1--5. IEEE.

\bibitem[Da~Silva, 1999]{da1999eeg}
Da~Silva, F.~L. (1999).
\newblock Eeg analysis: theory and practice.
\newblock {\em Electroencephalography: basic principles, clinical applications and related fields}, pages 1125--1159.

\bibitem[Dattola and La~Foresta, 2024]{dattola2024}
Dattola, S. and La~Foresta, F. (2024).
\newblock Application of electroencephalography (eeg) signal analysis in disease diagnosis.

\bibitem[De~Filippi et~al., 2021]{de2021classification}
De~Filippi, E., Wolter, M., Melo, B.~R., Tierra-Criollo, C.~J., Bortolini, T., Deco, G., and Moll, J. (2021).
\newblock Classification of complex emotions using eeg and virtual environment: Proof of concept and therapeutic implication.
\newblock {\em Frontiers in Human Neuroscience}, 15:711279.

\bibitem[Dhankhar et~al., 2022]{dhankhargwnet}
Dhankhar, N., Tiwari, R., Singh, T., and Buragohain, A. (2022).
\newblock Gwnet: Hierarchical and residual learning based 1d convolutional networks for gravitational wave detection.

\bibitem[Dustman et~al., 1999]{dustman1999life}
Dustman, R.~E., Shearer, D.~E., and Emmerson, R.~Y. (1999).
\newblock Life-span changes in eeg spectral amplitude, amplitude variability and mean frequency.
\newblock {\em Clinical neurophysiology}, 110(8):1399--1409.

\bibitem[Hamid et~al., 2020]{hamid2020temple}
Hamid, A., Gagliano, K., Rahman, S., Tulin, N., Tchiong, V., Obeid, I., and Picone, J. (2020).
\newblock The temple university artifact corpus: An annotated corpus of eeg artifacts.
\newblock In {\em 2020 IEEE Signal Processing in Medicine and Biology Symposium (SPMB)}, pages 1--4. IEEE.

\bibitem[Harati et~al., 2015]{harati2015improved}
Harati, A., Golmohammadi, M., Lopez, S., Obeid, I., and Picone, J. (2015).
\newblock Improved eeg event classification using differential energy.
\newblock In {\em 2015 IEEE Signal Processing in Medicine and Biology Symposium (SPMB)}, pages 1--4. IEEE.

\bibitem[Harati et~al., 2014]{harati2014tuh}
Harati, A., Lopez, S., Obeid, I., Picone, J., Jacobson, M., and Tobochnik, S. (2014).
\newblock The tuh eeg corpus: A big data resource for automated eeg interpretation.
\newblock In {\em 2014 IEEE signal processing in medicine and biology symposium (SPMB)}, pages 1--5. IEEE.

\bibitem[Hendrycks and Gimpel, 2016]{hendrycks2016gaussian}
Hendrycks, D. and Gimpel, K. (2016).
\newblock Gaussian error linear units (gelus).
\newblock {\em arXiv preprint arXiv:1606.08415}.

\bibitem[Jadhav et~al., 2022]{jadhav2022}
Jadhav, C., Kamble, P., Mundewadi, S., Jaiswal, N., Mali, S., Ranga, S., Suvvari, T.~K., and Rukadikar, A. (2022).
\newblock Clinical applications of eeg as an excellent tool for event related potentials in psychiatric and neurotic disorders.
\newblock {\em International Journal of Physiology, Pathophysiology and Pharmacology}, 14(2):73.

\bibitem[Jafari et~al., 2023]{jafari2023emotion}
Jafari, M., Shoeibi, A., Khodatars, M., Bagherzadeh, S., Shalbaf, A., Garc{\'\i}a, D.~L., Gorriz, J.~M., and Acharya, U.~R. (2023).
\newblock Emotion recognition in eeg signals using deep learning methods: A review.
\newblock {\em Computers in Biology and Medicine}, 165:107450.

\bibitem[Jiang et~al., 2025]{jiang2025neurolm}
Jiang, W., Wang, Y., liang Lu, B., and Li, D. (2025).
\newblock Neuro{LM}: A universal multi-task foundation model for bridging the gap between language and {EEG} signals.
\newblock In {\em The Thirteenth International Conference on Learning Representations}.

\bibitem[Jiang et~al., 2024]{labram2024}
Jiang, W.-B., Zhao, L.-M., and Lu, B.-L. (2024).
\newblock Large brain model for learning generic representations with tremendous eeg data in bci.
\newblock {\em arXiv preprint arXiv:2405.18765}.

\bibitem[Jing et~al., 2023]{jing2023development}
Jing, J., Ge, W., Hong, S., Fernandes, M.~B., Lin, Z., Yang, C., An, S., Struck, A.~F., Herlopian, A., Karakis, I., et~al. (2023).
\newblock Development of expert-level classification of seizures and rhythmic and periodic patterns during eeg interpretation.
\newblock {\em Neurology}, 100(17):e1750--e1762.

\bibitem[Kalita et~al., 2024]{kalita2024aneeg}
Kalita, B., Deb, N., and Das, D. (2024).
\newblock Aneeg: leveraging deep learning for effective artifact removal in eeg data.
\newblock {\em Scientific Reports}, 14(1):24234.

\bibitem[Koles, 1991]{koles1991quantitative}
Koles, Z.~J. (1991).
\newblock The quantitative extraction and topographic mapping of the abnormal components in the clinical eeg.
\newblock {\em Electroencephalography and clinical Neurophysiology}, 79(6):440--447.

\bibitem[Kondacs and Szab{\'o}, 1999]{kondacs1999long}
Kondacs, A. and Szab{\'o}, M. (1999).
\newblock Long-term intra-individual variability of the background eeg in normals.
\newblock {\em Clinical Neurophysiology}, 110(10):1708--1716.

\bibitem[Kostas et~al., 2021]{kostas2021bendr}
Kostas, D., Aroca-Ouellette, S., and Rudzicz, F. (2021).
\newblock Bendr: Using transformers and a contrastive self-supervised learning task to learn from massive amounts of eeg data.
\newblock {\em Frontiers in Human Neuroscience}, 15:653659.

\bibitem[Lawhern et~al., 2018]{lawhern2018eegnet}
Lawhern, V.~J., Solon, A.~J., Waytowich, N.~R., Gordon, S.~M., Hung, C.~P., and Lance, B.~J. (2018).
\newblock Eegnet: a compact convolutional neural network for eeg-based brain--computer interfaces.
\newblock {\em Journal of neural engineering}, 15(5):056013.

\bibitem[Lazarou et~al., 2018]{lazarou2018}
Lazarou, I., Nikolopoulos, S., Petrantonakis, P.~C., Kompatsiaris, I., and Tsolaki, M. (2018).
\newblock Eeg-based brain--computer interfaces for communication and rehabilitation of people with motor impairment: a novel approach of the 21 st century.
\newblock {\em Frontiers in human neuroscience}, 12:14.

\bibitem[Li et~al., 2022]{li2022motor}
Li, H., Ding, M., Zhang, R., and Xiu, C. (2022).
\newblock Motor imagery eeg classification algorithm based on cnn-lstm feature fusion network.
\newblock {\em Biomedical signal processing and control}, 72:103342.

\bibitem[Liao et~al., 2012]{liao2012gaming}
Liao, L.-D., Chen, C.-Y., Wang, I.-J., Chen, S.-F., Li, S.-Y., Chen, B.-W., Chang, J.-Y., and Lin, C.-T. (2012).
\newblock Gaming control using a wearable and wireless eeg-based brain-computer interface device with novel dry foam-based sensors.
\newblock {\em Journal of neuroengineering and rehabilitation}, 9:1--12.

\bibitem[Lopes~da Silva et~al., 2000]{lopes2000rhythms}
Lopes~da Silva, F., Pijn, J., Gorter, J., Van~Vliet, E., Daalman, E., and Blanes, W. (2000).
\newblock Rhythms of the brain: between randomness and determinism.
\newblock In {\em Chaos in Brain?}, pages 63--76. World Scientific.

\bibitem[Lopez et~al., 2015]{lopez2015automated}
Lopez, S., Suarez, G., Jungreis, D., Obeid, I., and Picone, J. (2015).
\newblock Automated identification of abnormal adult eegs.
\newblock In {\em 2015 IEEE signal processing in medicine and biology symposium (SPMB)}, pages 1--5. IEEE.

\bibitem[Loshchilov and Hutter, 2016]{cosinelr2016}
Loshchilov, I. and Hutter, F. (2016).
\newblock Sgdr: Stochastic gradient descent with warm restarts.
\newblock {\em arXiv preprint arXiv:1608.03983}.

\bibitem[Loshchilov and Hutter, 2017]{adamwoptimizer}
Loshchilov, I. and Hutter, F. (2017).
\newblock Decoupled weight decay regularization.
\newblock {\em arXiv preprint arXiv:1711.05101}.

\bibitem[Lotte et~al., 2015]{lotte2015}
Lotte, F., Bougrain, L., and Clerc, M. (2015).
\newblock Electroencephalography (eeg)-based brain-computer interfaces.
\newblock {\em Wiley encyclopedia of electrical and electronics engineering}, page~44.

\bibitem[M.~Shama et~al., 2023]{shama2023deepsoz}
M.~Shama, D., Jing, J., and Venkataraman, A. (2023).
\newblock Deepsoz: A robust deep model for joint temporal and spatial seizure onset localization from multichannel eeg data.
\newblock In {\em International Conference on Medical Image Computing and Computer-Assisted Intervention}, pages 184--194. Springer.

\bibitem[Ma et~al., 2023]{ma2023tsd}
Ma, Y., Liu, C., Ma, M.~S., Yang, Y., Truong, N.~D., Kothur, K., Nikpour, A., and Kavehei, O. (2023).
\newblock Tsd: Transformers for seizure detection.
\newblock {\em bioRxiv}, pages 2023--01.

\bibitem[Machado et~al., 2010]{machado2010eeg}
Machado, S., Ara{\'u}jo, F., Paes, F., Velasques, B., Cunha, M., Budde, H., Basile, L.~F., Anghinah, R., Arias-Carri{\'o}n, O., Cagy, M., et~al. (2010).
\newblock Eeg-based brain-computer interfaces: an overview of basic concepts and clinical applications in neurorehabilitation.
\newblock {\em Reviews in the Neurosciences}, 21(6):451--468.

\bibitem[Melman and Victor, 2016]{melman2016robust}
Melman, T. and Victor, J.~D. (2016).
\newblock Robust power spectral estimation for eeg data.
\newblock {\em Journal of neuroscience methods}, 268:14--22.

\bibitem[Mohammadi~Foumani et~al., 2024]{mohammadi2024eeg2rep}
Mohammadi~Foumani, N., Mackellar, G., Ghane, S., Irtza, S., Nguyen, N., and Salehi, M. (2024).
\newblock Eeg2rep: enhancing self-supervised eeg representation through informative masked inputs.
\newblock In {\em Proceedings of the 30th ACM SIGKDD Conference on Knowledge Discovery and Data Mining}, pages 5544--5555.

\bibitem[Noachtar and Rémi, 2009]{noachtar2022}
Noachtar, S. and Rémi, J. (2009).
\newblock The role of eeg in epilepsy: A critical review.
\newblock {\em Epilepsy \& Behavior}, 15(1):22--33.
\newblock Management of Epilepsy: Hope and Hurdles.

\bibitem[Nuwer, 1996]{nuwer1996quantitative}
Nuwer, M.~R. (1996).
\newblock Quantitative eeg analysis in clinical settings.
\newblock {\em Brain topography}, 8:201--208.

\bibitem[Panwar et~al., 2024]{panwar2024eeg}
Panwar, N., Pandey, V., and Roy, P.~P. (2024).
\newblock Eeg-cognet: A deep learning framework for cognitive state assessment using eeg brain connectivity.
\newblock {\em Biomedical Signal Processing and Control}, 98:106770.

\bibitem[Paszke et~al., 2019]{paszke2019pytorch}
Paszke, A., Gross, S., Massa, F., Lerer, A., Bradbury, J., Chanan, G., Killeen, T., Lin, Z., Gimelshein, N., Antiga, L., et~al. (2019).
\newblock Pytorch: An imperative style, high-performance deep learning library.
\newblock {\em Advances in neural information processing systems}, 32.

\bibitem[Peh et~al., 2022]{peh2022transformer}
Peh, W.~Y., Yao, Y., and Dauwels, J. (2022).
\newblock Transformer convolutional neural networks for automated artifact detection in scalp eeg.
\newblock In {\em 2022 44th Annual International Conference of the IEEE Engineering in Medicine \& Biology Society (EMBC)}, pages 3599--3602. IEEE.

\bibitem[Peksa and Mamchur, 2023]{peksa2023}
Peksa, J. and Mamchur, D. (2023).
\newblock State-of-the-art on brain-computer interface technology.
\newblock {\em Sensors}, 23(13):6001.

\bibitem[Percival and Walden, 1993]{percival1993spectral}
Percival, D.~B. and Walden, A.~T. (1993).
\newblock {\em Spectral analysis for physical applications}.
\newblock cambridge university press.

\bibitem[Press et~al., 2007]{press2007numerical}
Press, W.~H., Teukolsky, S.~A., Vetterling, W.~T., and Flannery, B.~P. (2007).
\newblock Numerical recipes third edition.

\bibitem[Roy et~al., 2018]{roy2018theory}
Roy, A., Vaswani, A., Neelakantan, A., and Parmar, N. (2018).
\newblock Theory and experiments on vector quantized autoencoders.

\bibitem[Roy et~al., 2019]{roy2019deep}
Roy, Y., Banville, H., Albuquerque, I., Gramfort, A., Falk, T.~H., and Faubert, J. (2019).
\newblock Deep learning-based electroencephalography analysis: a systematic review.
\newblock {\em Journal of neural engineering}, 16(5):051001.

\bibitem[Schreiter-Gasser et~al., 1994]{schreiter1994quantitative}
Schreiter-Gasser, U., Gasser, T., and Ziegler, P. (1994).
\newblock Quantitative eeg analysis in early onset alzheimer's disease: correlations with severity, clinical characteristics, visual eeg and cct.
\newblock {\em Electroencephalography and clinical Neurophysiology}, 90(4):267--272.

\bibitem[Shah et~al., 2018]{shah2018temple}
Shah, V., Von~Weltin, E., Lopez, S., McHugh, J.~R., Veloso, L., Golmohammadi, M., Obeid, I., and Picone, J. (2018).
\newblock The temple university hospital seizure detection corpus.
\newblock {\em Frontiers in neuroinformatics}, 12:83.

\bibitem[Shi et~al., 2024]{shi2024fome}
Shi, E., Zhao, K., Yuan, Q., Wang, J., Hu, H., Yu, S., and Zhang, S. (2024).
\newblock Fome: A foundation model for eeg using adaptive temporal-lateral attention scaling.
\newblock {\em arXiv preprint arXiv:2409.12454}.

\bibitem[Silipo et~al., 1998]{silipo1998dynamics}
Silipo, R., Deco, G., Vergassola, R., and Bartsch, H. (1998).
\newblock Dynamics extraction in multivariate biomedical time series.
\newblock {\em Biological Cybernetics}, 79(1):15--27.

\bibitem[Slepian and Pollak, 1961]{slepian1961prolate}
Slepian, D. and Pollak, H.~O. (1961).
\newblock Prolate spheroidal wave functions, fourier analysis and uncertainty—i.
\newblock {\em Bell System Technical Journal}, 40(1):43--63.

\bibitem[Song et~al., 2021]{song2021transformer}
Song, Y., Jia, X., Yang, L., and Xie, L. (2021).
\newblock Transformer-based spatial-temporal feature learning for eeg decoding.
\newblock {\em arXiv preprint arXiv:2106.11170}.

\bibitem[Song et~al., 2015]{song2015background}
Song, Y., Zang, D.-W., Jin, Y.-Y., Wang, Z.-J., Ni, H.-Y., Yin, J.-Z., and Ji, D.-X. (2015).
\newblock Background rhythm frequency and theta power of quantitative eeg analysis: predictive biomarkers for cognitive impairment post--cerebral infarcts.
\newblock {\em Clinical EEG and Neuroscience}, 46(2):142--146.

\bibitem[Song et~al., 2022]{song2022eeg}
Song, Y., Zheng, Q., Liu, B., and Gao, X. (2022).
\newblock Eeg conformer: Convolutional transformer for eeg decoding and visualization.
\newblock {\em IEEE Transactions on Neural Systems and Rehabilitation Engineering}, 31:710--719.

\bibitem[Stieger et~al., 2021]{stieger2021}
Stieger, J.~R., Engel, S.~A., Suma, D., and He, B. (2021).
\newblock Benefits of deep learning classification of continuous noninvasive brain-computer interface control.
\newblock {\em Journal of Neural Engineering}, 18(4):10.1088/1741--2552/ac0584.

\bibitem[Tang et~al., 2023]{tang2023modeling}
Tang, S., Dunnmon, J.~A., Liangqiong, Q., Saab, K.~K., Baykaner, T., Lee-Messer, C., and Rubin, D.~L. (2023).
\newblock Modeling multivariate biosignals with graph neural networks and structured state space models.
\newblock In {\em Conference on health, inference, and learning}, pages 50--71. PMLR.

\bibitem[Tang et~al., 2021a]{tang2021automated}
Tang, S., Dunnmon, J.~A., Saab, K., Zhang, X., Huang, Q., Dubost, F., Rubin, D.~L., and Lee-Messer, C. (2021a).
\newblock Automated seizure detection and seizure type classification from electroencephalography with a graph neural network and self-supervised pre-training.
\newblock {\em arXiv preprint arXiv:2104.08336}, 10.

\bibitem[Tang et~al., 2021b]{tang2021self}
Tang, S., Dunnmon, J.~A., Saab, K., Zhang, X., Huang, Q., Dubost, F., Rubin, D.~L., and Lee-Messer, C. (2021b).
\newblock Self-supervised graph neural networks for improved electroencephalographic seizure analysis.
\newblock {\em arXiv preprint arXiv:2104.08336}.

\bibitem[Tatum~IV, 2021]{tatum2021handbook}
Tatum~IV, W.~O. (2021).
\newblock {\em Handbook of EEG interpretation}.
\newblock Springer Publishing Company.

\bibitem[Thomson, 1982]{thomson1982spectrum}
Thomson, D. (1982).
\newblock Spectrum estimation and harmonic analysis.
\newblock {\em Proceedings of the IEEE}, 70(9):1055--1096.

\bibitem[Van Den~Oord et~al., 2017]{van2017neural}
Van Den~Oord, A., Vinyals, O., et~al. (2017).
\newblock Neural discrete representation learning.
\newblock {\em Advances in neural information processing systems}, 30.

\bibitem[van Vugt et~al., 2007]{van2007comparison}
van Vugt, M.~K., Sederberg, P.~B., and Kahana, M.~J. (2007).
\newblock Comparison of spectral analysis methods for characterizing brain oscillations.
\newblock {\em Journal of neuroscience methods}, 162(1-2):49--63.

\bibitem[Vaswani et~al., 2017]{vaswani2017attention}
Vaswani, A., Shazeer, N., Parmar, N., Uszkoreit, J., Jones, L., Gomez, A.~N., Kaiser, {\L}., and Polosukhin, I. (2017).
\newblock Attention is all you need.
\newblock {\em Advances in neural information processing systems}, 30.

\bibitem[Volkov, 2022]{volkov2022homology}
Volkov, I. (2022).
\newblock Homology-constrained vector quantization entropy regularizer.
\newblock {\em arXiv preprint arXiv:2211.14363}.

\bibitem[Wang et~al., 2023]{wang2023brainbert}
Wang, C., Subramaniam, V., Yaari, A.~U., Kreiman, G., Katz, B., Cases, I., and Barbu, A. (2023).
\newblock Brainbert: Self-supervised representation learning for intracranial recordings.
\newblock {\em arXiv preprint arXiv:2302.14367}.

\bibitem[Wang et~al., 2024a]{wang2024explain}
Wang, H., Yang, K., Zhang, J., Chen, T., and Song, L. (2024a).
\newblock Explain eeg-based end-to-end deep learning models in the frequency domain.
\newblock {\em arXiv preprint arXiv:2407.17983}.

\bibitem[Wang et~al., 2024b]{wang2024cbramod}
Wang, J., Zhao, S., Luo, Z., Zhou, Y., Jiang, H., Li, S., Li, T., and Pan, G. (2024b).
\newblock Cbramod: A criss-cross brain foundation model for eeg decoding.
\newblock {\em arXiv preprint arXiv:2412.07236}.

\bibitem[Wang et~al., 2024c]{wang2024graph}
Wang, L., Suzumura, T., and Kanezashi, H. (2024c).
\newblock Graph-enhanced eeg foundation model.
\newblock {\em arXiv preprint arXiv:2411.19507}.

\bibitem[Wu et~al., 2024]{wu2024neuro}
Wu, D., Li, S., Yang, J., and Sawan, M. (2024).
\newblock Neuro-bert: Rethinking masked autoencoding for self-supervised neurological pretraining.
\newblock {\em IEEE Journal of Biomedical and Health Informatics}.

\bibitem[Wu and He, 2018]{wu2018group}
Wu, Y. and He, K. (2018).
\newblock Group normalization.
\newblock In {\em Proceedings of the European conference on computer vision (ECCV)}, pages 3--19.

\bibitem[Yang et~al., 2023]{biot2023}
Yang, C., Westover, M.~B., and Sun, J. (2023).
\newblock Biot: Cross-data biosignal learning in the wild.
\newblock {\em arXiv preprint arXiv:2305.10351}.

\bibitem[Yang et~al., 2021a]{yang2021self}
Yang, C., Xiao, D., Westover, M.~B., and Sun, J. (2021a).
\newblock Self-supervised eeg representation learning for automatic sleep staging.
\newblock {\em arXiv preprint arXiv:2110.15278}.

\bibitem[Yang et~al., 2021b]{yang2021continental}
Yang, Y., Truong, N.~D., Maher, C., Nikpour, A., and Kavehei, O. (2021b).
\newblock Continental generalization of an ai system for clinical seizure recognition.
\newblock {\em arXiv preprint arXiv:2103.10900}.

\bibitem[Yuan et~al., 2024a]{yuan2024brainwave}
Yuan, Z., Shen, F., Li, M., Yu, Y., Tan, C., and Yang, Y. (2024a).
\newblock Brainwave: A brain signal foundation model for clinical applications.
\newblock {\em arXiv preprint arXiv:2402.10251}.

\bibitem[Yuan et~al., 2024b]{yuan2024brant}
Yuan, Z., Zhang, D., Chen, J., Gu, G., and Yang, Y. (2024b).
\newblock Brant-2: Foundation model for brain signals.
\newblock {\em arXiv e-prints}, pages arXiv--2402.

\bibitem[Zhang et~al., 2023a]{zhang2023spatial}
Zhang, B., Wei, D., Yan, G., Li, X., Su, Y., and Cai, H. (2023a).
\newblock Spatial--temporal eeg fusion based on neural network for major depressive disorder detection.
\newblock {\em Interdisciplinary Sciences: Computational Life Sciences}, 15(4):542--559.

\bibitem[Zhang et~al., 2023b]{zhang2023recent}
Zhang, J., Li, J., Huang, Z., Huang, D., Yu, H., and Li, Z. (2023b).
\newblock Recent progress in wearable brain--computer interface (bci) devices based on electroencephalogram (eeg) for medical applications: a review.
\newblock {\em Health data science}, 3:0096.

\end{thebibliography}

\clearpage
\appendix
\section{Appendix}
\subsection{Hyperparameters settings}

\begin{table}[H]
\small
\centering
\renewcommand{\arraystretch}{1.2} 
\caption{\textbf{Hyperparameters for EEG Tokenizer training}}
\label{tab:hps-tkn}
\begin{tabular}{ccc}
\hline
                                               & \textbf{Hyperparameters}   & \textbf{Values}                   \\ \hline
\multirow{7}{*}{Temporal Encoder}              & CNN Depth                  & 3                                 \\
                                               & Input channels             & \{1, 16, 16\}                     \\
                                               & Output channels            & \{16, 16, 16\}                    \\
                                               & Kernel size                & \{11, 3, 3\}                      \\
                                               & Input stride               & \{8, 1, 1\}                       \\
                                               & Input padding              & \{5, 1, 1\}                       \\
                                               & Activation function        & GELU                              \\ \hline
\multirow{3}{*}{Position \& Channel Embedding} & Num. time patches (T\textsubscript{W})     & 16                                \\
                                               & Num. channels patches (N\textsubscript{C}) & 16                                \\
                                               & Embedding dimension (D\textsubscript{E})   & 256                               \\ \hline
\multirow{5}{*}{Transformers}                  & encoder depth              & 12                                \\
                                               & decoder depth              & 3                                 \\
                                               & Attention heads            & 8                                 \\
                                               & MLP size                   & 1024                              \\
                                               & Activation Function        & GELU                              \\ \hline
\multirow{4}{*}{Quantizer}                     & Codebook vectors (C\textsubscript{V})      & 8192                              \\
                                               & Codebook dimension (C\textsubscript{D})    & 64                                \\
                                               & Beta                       & 0.3                               \\
                                               & Decay                      & 0.98                              \\ \hline
\multirow{12}{*}{Training}                     & Training size              & 4000 EEG hours                    \\
                                               & Batch size                 & 128                               \\
                                               & Peak learning rate         & 0.00032                           \\
                                               & Initial learning rate      & 0.000075                          \\
                                               & Minimal learning rate      & 1e-5                              \\
                                               & Learning Rate Scheduler    & Cosine                            \\
                                               & Optimizer                  & AdamW                             \\
                                               & AdamW Betas                & (0.9, 0.95)                       \\
                                               & Weight decay               & 0.01                              \\
                                               & Total epochs               & 100                               \\
                                               & Warmup epochs              & 10                                \\
                                               & Loss Function              & Commit Loss + Reconstruction Loss \\ \hline
\end{tabular}

\end{table}

\begin{table}[t]
\small
\centering
\renewcommand{\arraystretch}{1.2} 
\caption{\textbf{Hyperparameters for Masked Token Predictor training}}
\label{tab:hps-mtp}
\begin{tabular}{ccc}
\hline
                                               & \textbf{Hyperparameters}   & \textbf{Values} \\ \hline
\multirow{7}{*}{Temporal Encoder}              & CNN Depth                  & 3               \\
                                               & Input channels             & \{1, 16, 16\}   \\
                                               & Output channels            & \{16, 16, 16\}  \\
                                               & Kernel size                & \{11, 3, 3\}    \\
                                               & Input stride               & \{8, 1, 1\}     \\
                                               & Input padding              & \{5, 1, 1\}     \\
                                               & Activation function        & GELU            \\ \hline
\multirow{3}{*}{Position \& Channel Embedding} & Num. time patches (T\textsubscript{W})     & 16              \\
                                               & Num. channels patches (N\textsubscript{C}) & 16              \\
                                               & Embedding dimension (D\textsubscript{E})   & 256             \\ \hline
\multirow{4}{*}{Transformers}                  & encoder depth              & 12              \\
                                               & Attention heads            & 16              \\
                                               & MLP size                   & 1024            \\
                                               & Activation Function        & GELU            \\ \hline
\multirow{14}{*}{Training}                     & Training size              & 4000 EEG hours  \\
                                               & Batch size                 & 128             \\
                                               & Peak learning rate         & 0.0032          \\
                                               & Initial learning rate      & 8e-05           \\
                                               & Minimal learning rate      & 8e-05           \\
                                               & Learning Rate Scheduler    & Cosine          \\
                                               & Optimizer                  & AdamW           \\
                                               & AdamW Betas                & (0.9, 0.95)     \\
                                               & Weight decay               & 0.01            \\
                                               & Total epochs               & 30              \\
                                               & Warmup epochs              & 5               \\
                                               & Gradients clipping         & 1.0             \\
                                               & Mask ratio                 & 0.7             \\
                                               & Loss Function              & Cross-Entropy   \\ \hline
\end{tabular}
\end{table}

\begin{figure}[t]
\centering
\includegraphics[width=1.0\linewidth]{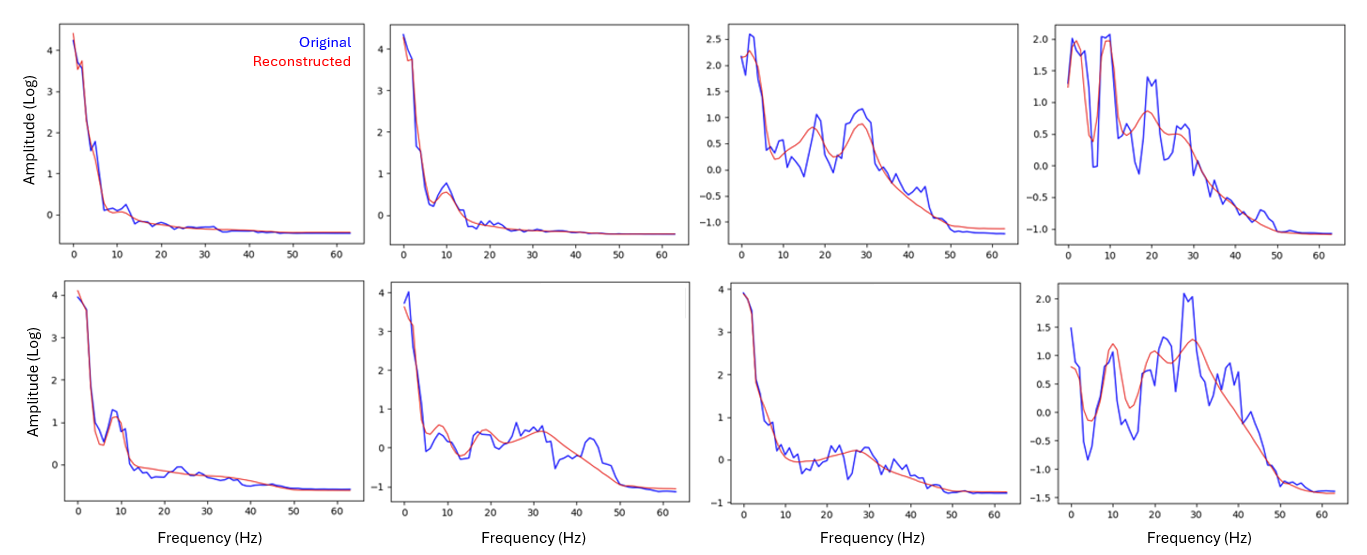}
\caption{\textbf{Original and reconstructed spectra.} Examples of the predicted EEG power distribution output by the pre-trained tokenizer decoder overlayed to the original multitaper estimate. Each subplot corresponds to the power spectrum distribution obtained from 1-second EEG window of one channel (i.e. one input patch). .}
\label{fig:tkn-spectra}
\end{figure}

\begin{figure}[t]
\centering
\includegraphics[width=1.0\linewidth]{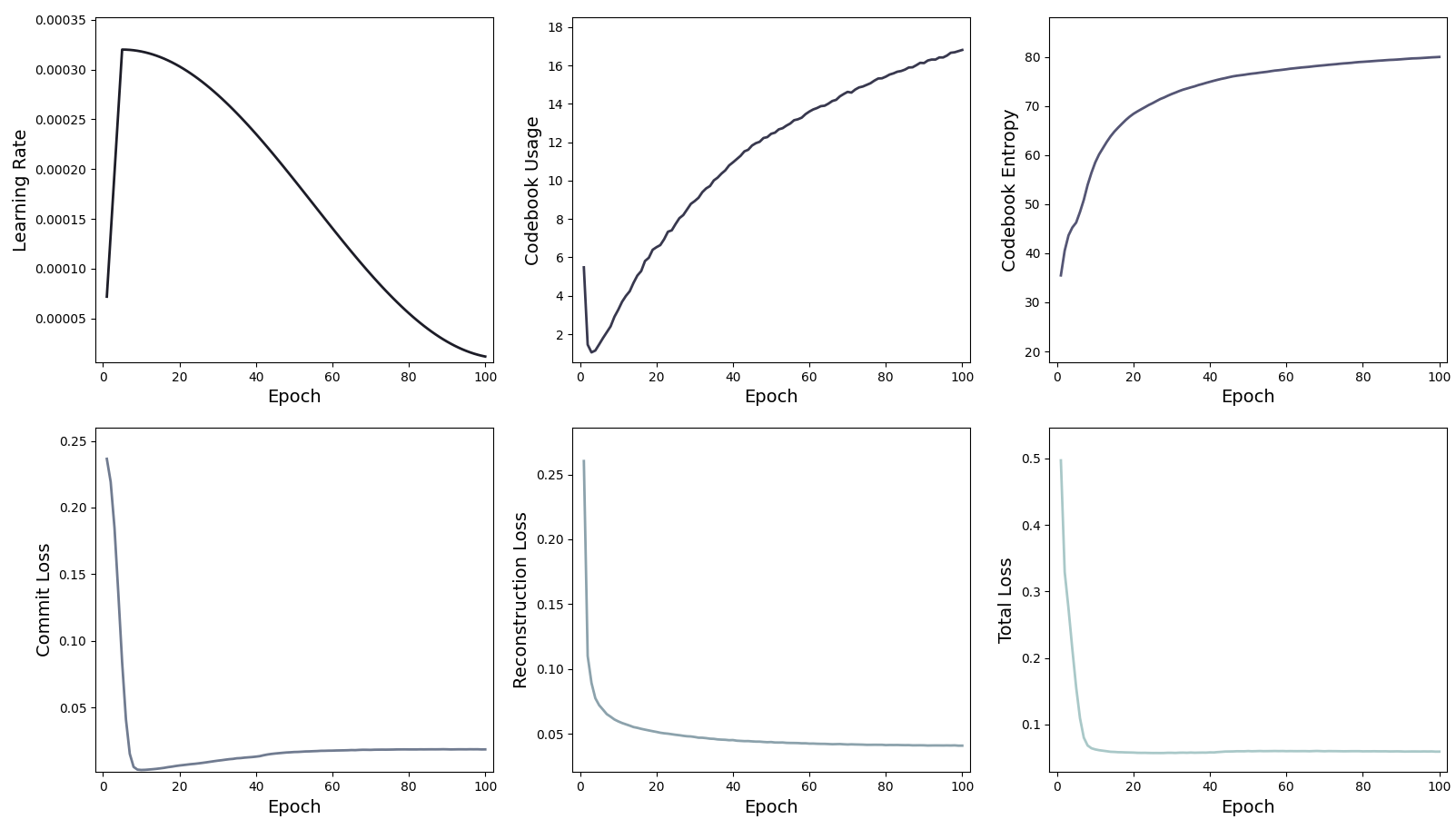}
\caption{\textbf{Tokenizer pre-training metrics.} From top-left to bottom-right: learning rate, codebook usage percentage, codebook normalized entropy, commit loss, reconstruction loss, total loss. Tokenizer backpropagation is based on total loss.}
\label{fig:tkn-pretrain-curves}
\end{figure}

\begin{figure}[t]
\centering
\includegraphics[width=1.0\linewidth]{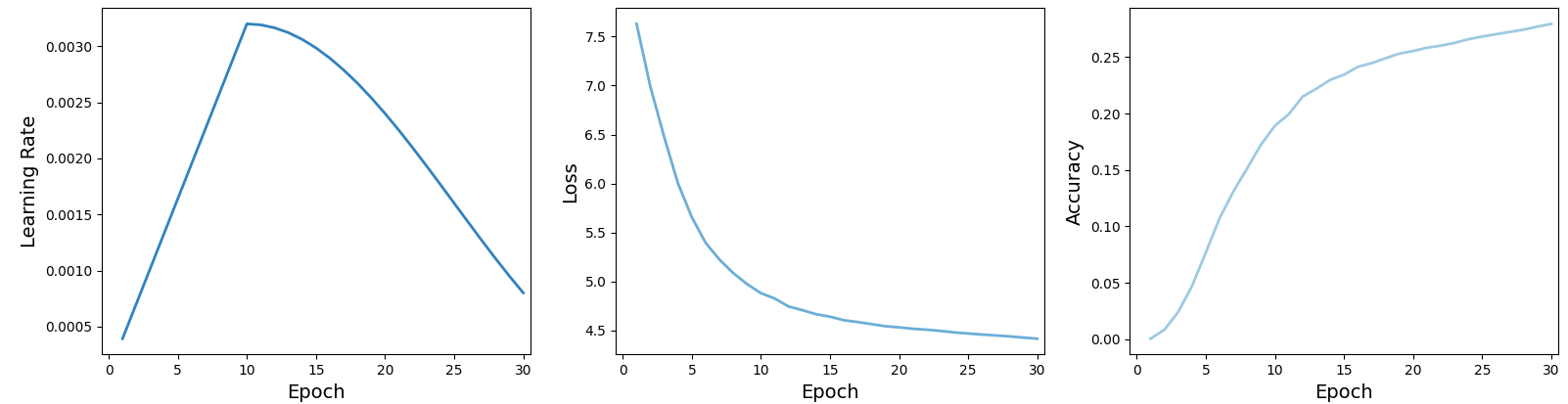}
\caption{\textbf{Masked token predictor pre-training metrics.} From left to right: learning rate, cross-entropy loss, accuracy in predicting the correct codebook indices from both masked and unmasked input patches.}
\label{fig:mtp-pretrain-curves}
\end{figure}

\end{document}